\documentclass[aps,prd,twocolumn,superscriptaddress,showpacs,footinbib,preprintnumbers]{revtex4-1}
\usepackage{hyperref}
\usepackage{slashed}
\usepackage{bm}
\usepackage{graphicx} 
\usepackage{enumerate}
\usepackage{amsfonts,amsmath,amssymb,bm,bbm}
\usepackage{color}
\usepackage{epsfig, subfigure}
\usepackage{setspace}  
\usepackage{booktabs, tabularx} 
\usepackage{multirow}

\hypersetup{ 
    pdfnewwindow=true,      
    colorlinks=true,       
    linkcolor=blue,          
    citecolor=blue,        
    filecolor=blue,      
    urlcolor=blue        
}  

 \definecolor{pink}{rgb}{0.7,0,0.4}

\newcommand{\be}{\begin{equation}}  
\newcommand{\ee}{\end{equation}}
\newcommand{\bea}{\begin{eqnarray}}  
\newcommand{\eea}{\end{eqnarray}}

\begin{document}

\title{Double-Regge Exchange Limit for the $\gamma p\rightarrow K^+K^-p$ Reaction}

\author{M.~Shi}\email{shimeng1031@pku.edu.cn}
\affiliation{Department of Physics, Peking University, Beijing 100871, China}
\affiliation{Theory Center, Thomas Jefferson National Accelerator Facility, Newport News, VA 23606}

\author{I.V. Danilkin}
\affiliation{Theory Center, Thomas Jefferson National Accelerator Facility, Newport News, VA 23606}

\author{C. Fern\'andez-Ram\'{\i}rez}
\affiliation{Theory Center, Thomas Jefferson National Accelerator Facility, Newport News, VA 23606}

\author{V. Mathieu}
\affiliation{Center for Exploration of Energy and Matter, Indiana University, Bloomington, IN 47403}
\affiliation{Physics Department, Indiana University, Bloomington, IN 47405}

\author{M. R. Pennington}
\affiliation{Theory Center, Thomas Jefferson National Accelerator Facility, Newport News, VA 23606}

\author{D. Schott}
\affiliation{Department of Physics, The George Washington University, Washington,  DC 20052}

\author{A. P. Szczepaniak}
\affiliation{Theory Center, Thomas Jefferson National Accelerator Facility, Newport News, VA 23606}
\affiliation{Center for Exploration of Energy and Matter, Indiana University, Bloomington, IN 47403}
\affiliation{Physics Department, Indiana University, Bloomington, IN 47405}

\collaboration{Joint Physics Analysis Center}
\preprint{JLAB-THY-14-1984}

\date{\today}

\begin{abstract}
We apply the generalized Veneziano model ($B_5$ model) in the double-Regge exchange limit to the $\gamma p \to K^+ K^- p$ reaction.
Four different cases defined by the possible combinations of the signature factors of leading Regge exchanges 
(($K^*,a_2/f_2$), ($K^*,\rho/\omega$), ($K_2^*,a_2/f_2$), and ($K_2^*,\rho/\omega$)) have been simulated through the Monte Carlo method.
Suitable event candidates for the double-Regge exchange high-energy limit were selected employing \textit{Van Hove} 
plots as a better alternative to kinematical cuts in the $K^+ K^- p$ Dalitz plot. 
In this way we predict and analyze the double-Regge contribution to the $K^+ K^- p$ Dalitz plot, which constitutes one of the major backgrounds in the search for
 strangeonia, hybrids and exotics using $\gamma p\rightarrow K^+K^-p$ reaction.
We expect that data currently under analysis, and that to come in the future, will allow
verification of the double-Regge behavior and a better assessment of this component of the amplitude.
\end{abstract}

\maketitle

\section{Introduction}
\label{sec:introduction}
 
A host of new experiments  dedicated to precision studies of the hadron spectrum will begin 
operations in the near future. 
 These will complement and extend the reach of recently completed and other ongoing experiments 
 that, among other discoveries,  found intriguing structures in the hadron spectrum
 \cite{Belle3872,Belle4430,LHCb443201,LHCb44302,BelleZb,BES3900,Naturequartet}. Existence of 
 these structures demonstrates that the hadron spectrum is far more complex than predicted by the 
 valence quark model \cite{Gellmann}. Nevertheless, it remains to be seen if these structures are to be 
 associated with new resonances. 
This is because identification of new states requires detailed understanding of reaction dynamics. 
The tools that enable us to constrain and interpret reaction amplitudes are based on principles of the 
 S-matrix theory,  which include analyticity, crossing relations and unitarity. In practice, rigorous  
implementation of these principles is impossible. It would require knowledge of an infinite number of 
amplitudes describing all coupled channels for all reactions related by crossing.  Nevertheless, for a 
given reaction it is  possible to kinematically isolate regions where specific processes dominate and use 
analyticity to constrain amplitudes in other regions of interest, \textit{e.g.} correlate amplitude 
parametrization in the  low- and the high-energy regions using finite energy sum rules~\cite{FESR}. 
    
It follows from S-matrix principles that in relativistic scattering resonance formation in the direct 
channel is dual to Regge exchanges, \textit{aka} Reggeons, in the cross-channels. Leading Reggeons in the 
cross-channel determine the high energy behavior of the direct channel.  Thus, because of analyticity,  
contributions from resonances at low-energies are smoothly connected 
with Reggeon contributions at the higher energies. Therefore identification of resonances has to be made 
simultaneously with studies of the high-energy behavior and cross-channel Regge exchanges 
\cite{Collins}.
   
       A class of models that incorporates resonance-Regge duality has been extensively studied in the past \cite{Chew-Pignotti,Barger-Durand,Allesandrini-Amati-Squire,Collins-Ross-Squires}. These dual models are based on an extension of the Veneziano \cite{G.Veneziano_1} approach for amplitudes connecting four external particles to reactions with an arbitrary number, $N$, of external particles
       \cite{ChanN,Goebel-Sakita,Koba-Nielsen}. The simplest, so-called  $B_N$ dual model satisfies crossing and resonance-Regge duality for linear trajectories. Even though the $B_N$ model lacks proper unitarity, which would require non-linear trajectories, it is expected to provide a reasonable description of reaction amplitudes when averaged over resonance widths. Various extensions that enable unitarity, and as a consequence implement complex trajectories 
        ~\cite{complex_trajectories1,complex_trajectories2}, have been proposed but they lack the simplicity of the original formulation \cite{b5-application1,b5-application2,b5-application3,b5-application4}.

In this paper we apply the $B_5$ model \cite{B5first1,B5first,pokorski-bialas} in the double-Regge exchange limit (DRL) to the reaction $\gamma p \to K^+ K^- p$. 
The analysis of this reaction is currently underway based on the data collected by the CLAS collaboration at JLab  
 using the highest photon energy beam, $E_\gamma \le 5.5 \mbox{ GeV}$, delivered to date at CEBAF to CLAS.
 The $K\bar K$ spectrum produced in photon  dissociation is expected to be dominated by vector resonances, but higher-spin  
  states are also possible. At present, however, there is little evidence for $K\bar K$ decay modes of higher mass meson resonances~\cite{PDG2014}, suggesting resonance signals in the $K\bar K$ channel of $\gamma p \to K^+ K^- p$ may be weak. 
 This makes studies of non-resonant processes even more relevant. The Regge/Pomeron exchange is the dominant  process in the kinematical domain where all sub-channel invariants are large. According to the hypothesis of two-component duality~\cite{HHarari}, cross-channel Regge exchanges are dual to direct channel resonances. Thus analysis of the $K\bar K$ spectrum in photon-production will benefit from understanding of the DRL of this reaction.   
  
 The rest of the paper is organized as follows. In Sec.\,\ref{sec:DualModel}, we discuss properties of the $B_N$ dual models  focusing on $B_4$ (the Veneziano model)  and $B_5$, which are of relevance to the process of interest here. 
  In Sec.\,\ref{B5amp}, we describe the double-Regge limit of the $B_5$ amplitude. 
Once the structure of the dual amplitude in the double-Regge limit is obtained, in Sec.\,\ref{sec:Spin Structure}, we introduce appropriate modifications related to the presence of external particles with non-zero spin that make the $B_5$ model more suitable for analysis of kaon-pair photoproduction. There we  also present results of our numerical analysis. The summary and outlook are given in Sec.\,\ref{sec:conclusion}.
 
\section{Dual Amplitude Model}
\label{sec:DualModel}
The Veneziano model~\cite{G.Veneziano_1} describes an amplitude of four external scalar particles.  As such, it is a function of the three Mandelstam variables $s,t,u$, which are related by
\begin{equation}  
s + t + u = \Sigma, \label{mand}
\end{equation} 
where $\Sigma$  is the sum of squares of masses of the external particles. In the following, all dimensional quantities are measured in units  of $\mbox{ GeV}$. The building block of the Veneziano model is the $B_4$  amplitude. It is a function of two variables, {\it e.g.} $s$ and $t$. For four particles, any pair of Mandelstam variables  corresponds to invariant mass squared in two overlapping channels. The $B_4$ amplitude has Regge behavior in each channel and for linear trajectories, $\alpha(x)=\alpha_0+\alpha' x$, exhibits duality between resonances in one channel and Reggeons in the overlapping channel.  Assuming, for simplicity that the reaction is $s\leftrightarrow t$ symmetric, {\it i.e.} resonance/Regge trajectories in the $s$ and $t$ channels are identical, a $B_4$ amplitude can be written as, 
\begin{equation}
\begin{split}
B_4(s,t)=&\frac{\Gamma(-\alpha(s))\Gamma(-\alpha(t))}{\Gamma(-\alpha(s)-\alpha(t))} \\
= & \sum_{n=0}^{\infty}\frac{\beta_n(t)}{n-\alpha(s)} 
=\sum_{n=0}^{\infty}\frac{\beta_n(s)}{n-\alpha(t)}.  \\
\end{split}
\label{b4} 
\end{equation}
For linear trajectories the residue function, 
\begin{equation} 
\beta_n(x) = \frac{\Gamma(-\alpha(x))}{\Gamma(-n-\alpha(x))},\label{beta}
\end{equation}
 is a polynomial in $x$ of order $n$.   The two alternative forms in Eq.~(\ref{b4}) represent the amplitudes for spinless particle scattering in terms of an infinite series of narrow resonances in either the $s$ or the $t$ channel. The solution of the equation $\alpha(m_R^2) = n$ gives the mass $m_R$  of the resonances. In the model, at given $n$ there are $n+1$  
  degenerate resonances with spins ranging from $0$ to $n$.   The couplings of these resonances to the 
 external particles are computed by expanding the residue function $\beta_n(x)$ in terms of Legendre polynomials. In the Veneziano model 
  couplings are fixed and determined by the ratio of $\Gamma$ functions in Eq.~(\ref{beta}). A model with adjustable couplings may be  obtained by taking combinations of the $B_4$'s with different parameters, {\it i.e.} trajectory intercept $\alpha_0$, slope $\alpha'$ and the overall normalization~\cite{Sivers,AdamMichael}.

The asymptotic behavior of the amplitude in Eq.~(\ref{b4}) in the limit when one of the channel variables, {\it e.g.} $s$ is taken to infinity   $s\rightarrow \infty$, follows from the Stirling's formula (shown in Eq.\,(\ref{Stirlingformula})) and is given by 
\begin{equation}
 B_4(s,t)\rightarrow (-\alpha(s))^{\alpha(t)}\Gamma(-\alpha(t)), 
\end{equation} 
which, except for the signature factor, is the behavior expected for $t$-channel exchange of a Regge trajectory. 
Regge signature factors are recovered by taking appropriate linear combinations of the $B_4$ amplitudes with different channel variables as arguments. For example, at fixed-$t$, the amplitude is symmetric under $s \leftrightarrow u$ crossing, contains only signature-even $t$-channel Reggeons and corresponds to a combination $B_4(s,t) + B_4(u,t)$. The leading behavior in the $s\rightarrow \infty$, limit  is then given by 
\begin{equation}
\begin{split}
B_4(s,t)+&B_4(u,t) \rightarrow \\
& \left[ (-\alpha(s))^{\alpha(t)} + (-\alpha(u))^{\alpha(t)} \right]  \Gamma(-\alpha(t)).
\end{split} \label{b4s} 
\end{equation}

\begin{figure}
\rotatebox{0}{\scalebox{0.42}[0.42]{\includegraphics{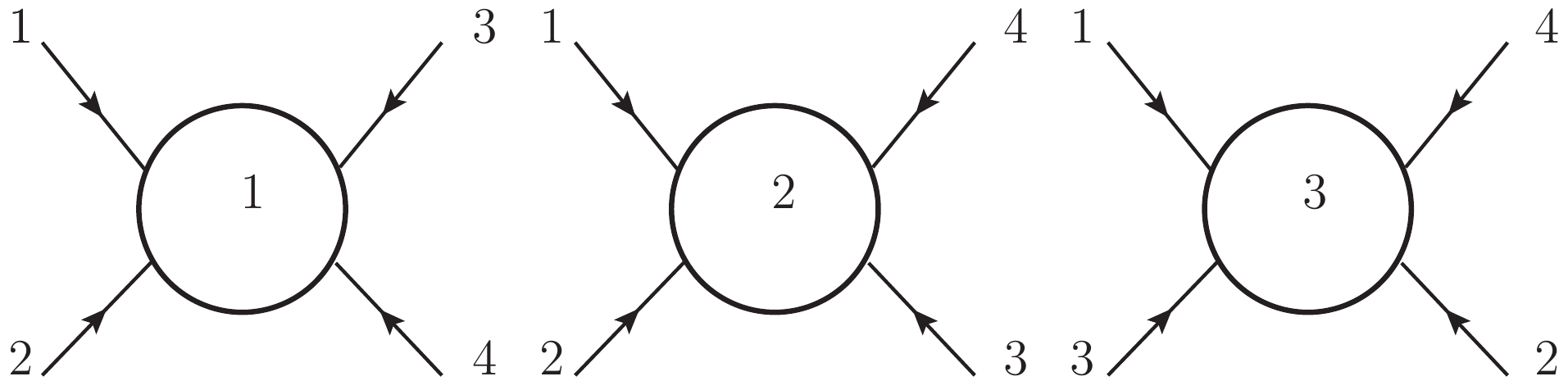}}}
\caption{Diagramatic representation of the three independent $B_4$ amplitudes. The Mandelstam invariants, $s_{ij} = s_{ji} =  (p_i + p_j)^2$ of the neighboring overlapping pairs are the arguments of the amplitude. The amplitudes, from left to right are given by, $B_4(s_{12},s_{13})$, $B_4(s_{12},s_{14})$ and $B_4(s_{13},s_{14})$. } 
\label{fig:threeB4}
\end{figure}

So far we have assumed that all trajectory functions are linear. If one also  assumes a common slope then it follows from Eq.~(\ref{mand}) that, 
\bea
\alpha(s)+\alpha(t)+\alpha(u)=\mbox{const.} 
\eea
and  the leading behavior at large-$s$  of the combination in Eq.~(\ref{b4s}) reduces to 
\bea \label{r4} 
B_4(s,t)+B_4(u,t)\rightarrow (-\alpha(s))^{\alpha(t)}\xi(t)  \Gamma(-\alpha(t))
\eea
where $\xi(t) = 1 + \tau e^{i\pi\alpha(t)}$ with $\tau = +1$ is the proper signature factor for the spin-even $t$-channel Regge  exchange. Thus Eq.~(\ref{r4}) is consistent with the expectations from Regge theory for the contribution of the leading Regge pole.   At fixed $t$ and large and positive $s$, $s\to \infty$, $u \to -\infty$ the amplitude without $t$-channel poles, {\it i.e.} $B_4(u,s)$ behaves as 
\begin{equation} 
B_4(u,s) \to e^{i\pi \alpha(s)} \sim e^{- i \pi \alpha(u)}  \label{bus} 
\end{equation} 
When the simple linear trajectory is replaced with a realistic one, which has a positive imaginary part that grows as $s\to \infty$, the $B_4(u,s)$ amplitude becomes exponentially suppressed. Thus, as expected from Regge theory, in the $B_4$ dual model, out of the three possible diagrams shown in Fig.~\ref{fig:threeB4}, only two contribute in the Regge limit, $s\to \infty$ and $t$-fixed. 

\begin{figure}
\rotatebox{0}{\scalebox{0.35}[0.35]{\includegraphics{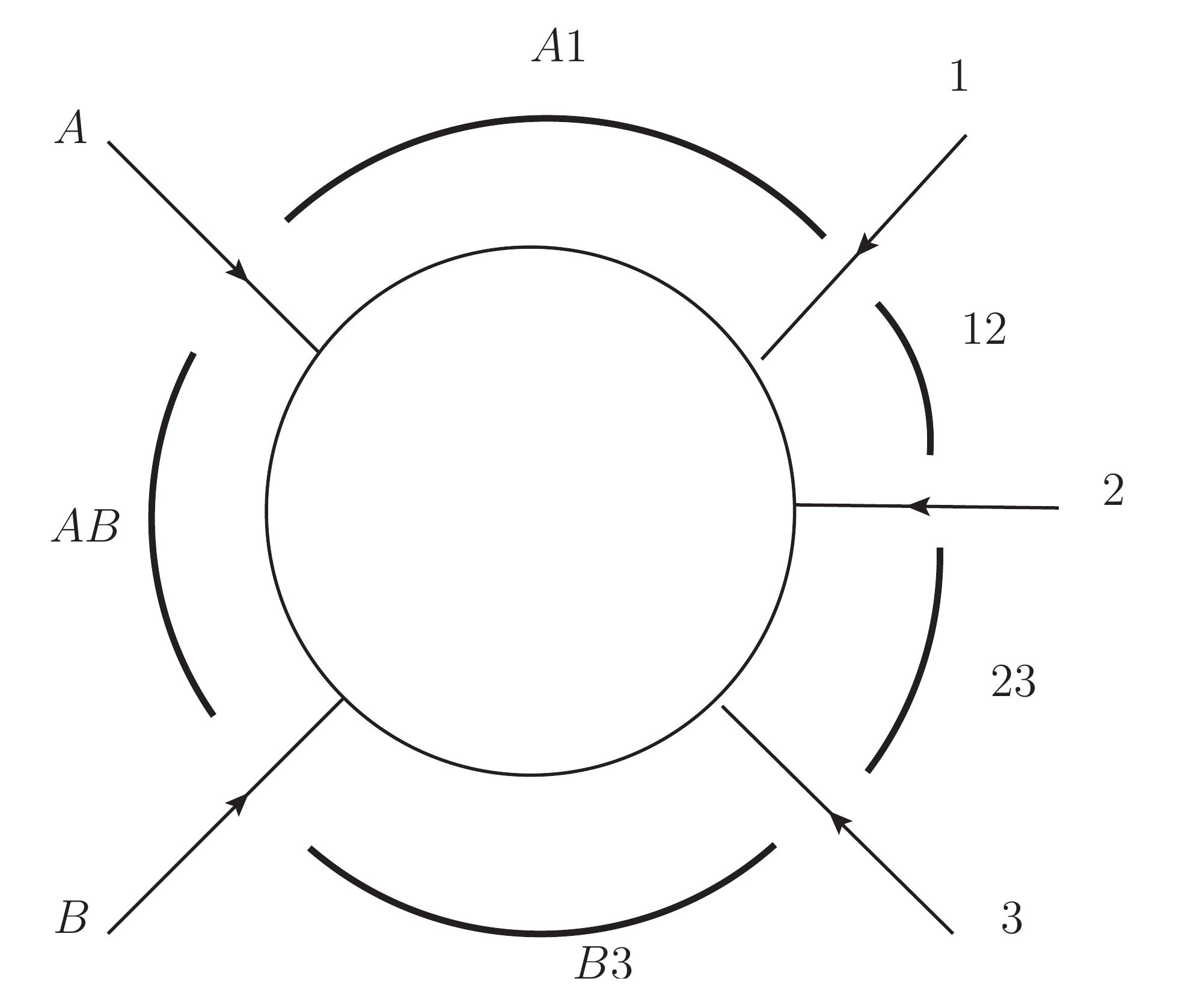}}}
\caption{Representation of the $B_5(s_{AB},s_{A1},s_{12},s_{23},s_{B3})$ amplitude and its kinematics. 
As in Fig.~\ref{fig:threeB4}, the Mandelstam invariants, $s_{ij} = s_{ji} =  (p_i + p_j)^2$ 
of the neighboring overlapping pairs are the arguments of the amplitude. } 
\label{fig:B5round}
\end{figure}

The leading meson trajectory  is approximately equal to 
$\alpha(s) = 0.5 + s$. Therefore, the first pole in the amplitude of Eq.~(\ref{b4}), which corresponds to $\alpha=0$, would correspond to a resonance of spin-0 and negative mass squared $m_R^2 = -0.5$. A spurious pole like this is easily removed by replacing $-\alpha$ by $1-\alpha$ in the arguments of the $\Gamma$ functions. With such a shift, the first pole in the amplitude corresponds to a spin-1 or spin-0 resonance 
 with mass $m_R^2 = 0.5\mbox{ GeV}^2$.   
 
 The amplitude $B_4(x,y)$ in Eq.~(\ref{b4}) is identical to the Euler's Beta-function. Thus $B_4$ can also be written using an integral representation, which is defined for  $\alpha(s),\alpha(t) < 0$, to be
 \bea
B_4(s,t)=\int_0^1du\, u^{-\alpha(s)-1}(1-u)^{-\alpha(t)-1}.
\eea 
 For other values of $\alpha(s)$ and $\alpha(t)$, the amplitude is obtained from Eq.~(\ref{b4}), {\it i.e.} by analytical continuation. The integral representation provides the basis for generalization of the Veneziano amplitude to an arbitrary number of external particles. In particular the $B_5$ amplitude can be written as, 
\begin{widetext}
\begin{equation}
B_5(s_{AB},s_{A1},s_{12},s_{23},s_{B3}) =\int_0^1 dt \: \int_0^1 du\, t^{-\alpha_{12}-1}u^{-\alpha_{23}-1}
\left(1-t \right)^{-\alpha_{A1}-1}
\left(1-u\right)^{-\alpha_{B3}-1}\left(1-tu\right)^{-\alpha_{AB}+\alpha_{12}+\alpha_{23}}  \label{b5i} 
\end{equation}
\end{widetext}
where $\alpha_{ij}=\alpha_{0,ij}+\alpha'_{ij}s_{ij}$ and $s_{ij} = (p_i + p_j)^2$ are  the channel variables. We 
 adopt the labeling convention from  \cite{pokorski-bialas} with all particle momenta, $p_i$ taken as incoming, {\it cf.} 
Fig.~\ref{fig:B5round} and  $i,j=A, B, 1, 2, 3$. Using the bar to represent an antiparticle, the reaction $ A + B \to \bar 1 + \bar 2 + \bar 3$ corresponds the physical channel of the reaction of interest, {\it i.e.} $\gamma  =  A$, $p(\mbox{target})  =  B$, $K^+ = \bar 1$, $K^- =\bar 2 $, $p(\mbox{recoil}) = \bar 3$. The reaction amplitude involving five particles depends on five independent kinematical variables,  which we choose as the consecutive two-body channel invariants, $s_{ij}$ as shown in Fig.~\ref{fig:B5round}.   It follows from the integral representation in Eq.~(\ref{b5i}) that $B_5$ is symmetric under cyclic permutation and the reflection of the arguments.  The integral representation in Eq.~(\ref{b5i}) is valid when all trajectories are negative, $\alpha_{ij} < 0$,  which is outside the physical region of the reaction of interest. The amplitude in the physical region is obtained by analytic continuation. 
As in the case of the $B_4$ model, the analytical continuation of Eq.~(\ref{b5i}) is performed once the integral is represented  in terms of analytical functions. In particular it can be expressed in terms of Euler Beta-functions,  ($B_4$ amplitudes) and a generalized hypergeometric function of unit argument, 
\begin{widetext} 
\begin{equation}
\begin{split}
B_5(s_{AB},s_{A1},s_{12},s_{23},s_{B3})= & \: B_4(-\alpha_{12},-\alpha_{A1})\: B_4(-\alpha_{23},-\alpha_{B3})  \\
& \times \:  _3F_2(\alpha_{AB} 
-\alpha_{12}-\alpha_{23},-\alpha_{A1},-\alpha_{B3};-\alpha_{12}-\alpha_{A1},-\alpha_{23}-\alpha_{B3}). 
\end{split}
\label{eq:B5}
\end{equation}
\end{widetext} 
The relevant properties of generalized hypergeometric functions are given in the Appendix.  The series expression for the hypergeometric function of unit argument, $_3F_2(a,b,c;d,e)$ is convergent provided  $\text{Re}(a+b+c-d-e)<0$,  which implies Eq.\,(\ref{eq:B5}) is well defined for $\alpha_{AB}<0, (s_{AB}<0)$ but ill-defined for $\alpha_{AB}>0,  (s_{AB}>0)$. It is the latter that corresponds to the physical region of $\gamma p \to K^+ K^- p$. The symmetry properties of the $B_5$, 
{\it e.g.}
\begin{widetext}
\begin{equation}
\begin{split}
B_5(s_{AB},s_{A1},s_{12},s_{23},s_{B3}) = &\: B_5(s_{A1},s_{12},s_{23},s_{B3},s_{AB}) \\
 =  B_4(-\alpha_{12},-\alpha_{23})& B_4(-\alpha_{B3},-\alpha_{AB}) 
\, _3F_2(\alpha_{A1}-\alpha_{23}-\alpha_{B3},-\alpha_{12},-\alpha_{AB};-\alpha_{12}-\alpha_{23},-\alpha_{B3}-\alpha_{AB}).
\end{split}
\label{B5:identity}
\end{equation}
\end{widetext}
 enables the analytical continuation  of Eq.\,(\ref{eq:B5}) to the physical region of $ A + B \to \bar 1 + \bar 2 + \bar 3$.
Alternatively, one can use the relations between hypergeometric functions given in the Appendix ({\it cf.} Eq.\,(\ref{eq:HYPrelation2})) \cite{pokorski-bialas} to continue to the region ($s_{AB}>0$). Both continuations   
 result in the same amplitude \cite{numerical_code}.

\section{Double-Regge limit of $B_5$ amplitude}
\subsection{Scalar Amplitudes}
\label{B5amp}

\begin{figure}
\rotatebox{0}{\scalebox{0.42}[0.42]{\includegraphics{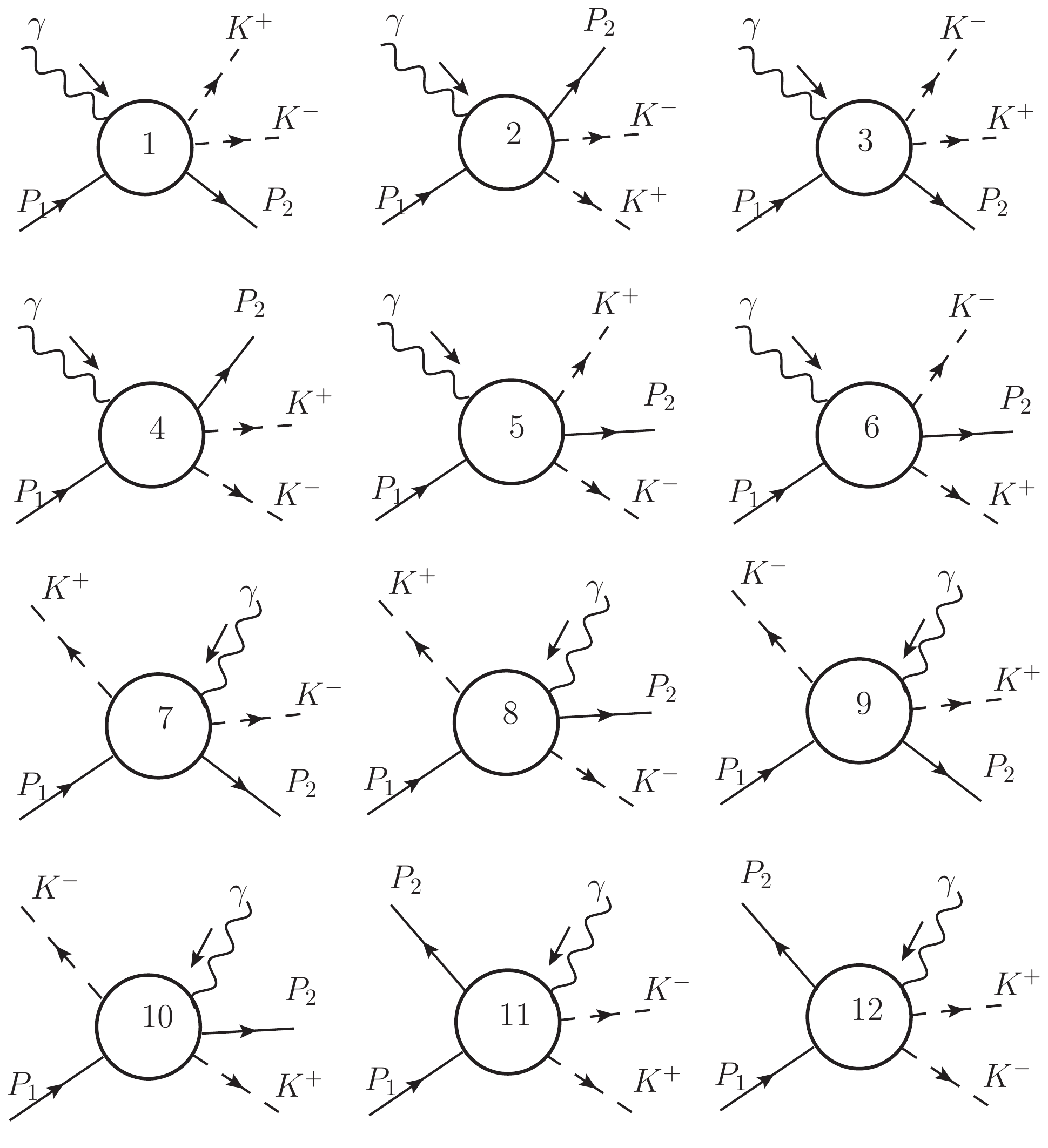}}}
\caption{Twelve diagrams involved in the reaction $\gamma p\rightarrow K^+K^-p$. 
The diagrams, labeled as $B_5(i)$ and distinguished by the channel invariants they depend upon
 are discussed in Sec. \ref{B5amp} and Fig.~\ref{fig:B5round} caption. } 
\label{fig:All diagrams}
\end{figure}

Taking into account symmetries implied by  Eq.~(\ref{b5i}) out of $5$! permutations of the external particles there are only 12 independent amplitudes. These are the equivalent to the 3 independent $B_4(x,y)$, $x,y=s,t,u$ amplitudes of the Veneziano model shown in Fig.~\ref{fig:threeB4}. The twelve amplitudes for the reaction $\gamma p\rightarrow K^+K^-p$, denoted by $B_5(i)$, $i=1\dots 12$,  are depicted in Fig.~{\ref{fig:All diagrams}}. The first six diagrams have the photon and incident proton next to each other and the other six diagrams have one particle between the incident photon and proton.
 Before imposing additional symmetry constraints, {\it e.g.} Bose symmetry, the 
most general 5-point amplitude is given by a linear combination of the 12 independent $B_5$ amplitudes.

For a $2-$to$-3$ reaction, the double-Regge limit corresponds to large values of the channel energies and small momentum transfers, 
\bea\label{doubleRegge:condition}
&&s_{AB}, s_{12}, s_{23} \rightarrow \infty,  \frac{s_{12}s_{23}}{s_{AB}} = \mbox{ fixed} \nonumber\\
 &&\hspace{1cm}\frac{t_{A1}}{s_{AB}}, \frac{t_{B3}}{s_{AB}} \to 0,
 \eea
 where $t_{A1}=(p_A-p_1)^2$ and $t_{B3}=(p_B-p_3)^2$.
To compute the double-Regge limit of the $B_5(1)$ amplitude we use the relations (\ref{eq:HYPrelation1}) and (\ref{eq:HYPrelation2}) in the Appendix to obtain 
\bea\label{separateB5}
&&\hspace{-0.1cm}B_5=B_4(-\alpha_{AB},-\alpha_{A1})B_4(-\alpha_{23},-\alpha_{B3}+\alpha_{A1})\nonumber\\
&&\hspace{1.2cm}\times\,_3F_2(-\alpha_{A1},-\alpha_{23},-\alpha_{12}-\alpha_{A1}+\alpha_{B3};\nonumber\\
&&\hspace{2.2cm}1-\alpha_{A1}+\alpha_{B3},-\alpha_{AB}-\alpha_{A1})\nonumber\\
&&\hspace{0.8cm}+B_4(-\alpha_{AB},-\alpha_{B3})B_4(-\alpha_{12},-\alpha_{A1}+\alpha_{B3})\nonumber\\
&&\hspace{1.2cm}\times\,_3F_2(-\alpha_{B3},-\alpha_{12},-\alpha_{23}-\alpha_{B3}+\alpha_{A1};\nonumber\\
&&\hspace{2.2cm}1-\alpha_{B3}+\alpha_{A1},-\alpha_{AB}-\alpha_{B3}). 
\eea
In the above expression three out of the five arguments in the two hypergeometric functions are large. Using the relation 
  \bea\label{eq:highenergyHPY}
_3F_2(a,\mu,\nu;b,\lambda;z)\rightarrow \,_1F_1(a,b;\frac{\mu\nu z}{\lambda}),
\eea
valid in the limit $\lambda,\,\mu,\,\nu\rightarrow \infty$, Eq.\,(\ref{separateB5}) reduces to 
\bea\label{eq:highenergyB5}
B_5(1)\rightarrow 
&&B_4(-\alpha_{AB},-\alpha_{A1})B_4(-\alpha_{23},-\alpha_{B3}+\alpha_{A1})\nonumber\\
&&\hspace{0.3cm}\times_1F_1(-\alpha_{A1};1-\alpha_{A1}+\alpha_{B3};-\eta)\nonumber\\
&&+B_4(-\alpha_{AB},-\alpha_{B3})B_4(-\alpha_{12},-\alpha_{A1}+\alpha_{B3})\nonumber\\
&&\hspace{0.3cm}\times_1F_1(-\alpha_{B3};1-\alpha_{B3}+\alpha_{A1};-\eta),
\eea
where $\eta=\alpha_{12}\alpha_{23}/\alpha_{AB}$. Further simplification is obtained when in the two limits, 
 $\eta\rightarrow 0$ or $\eta\rightarrow \infty$. In the former (case a)  $_1F_1(a;b;-\eta)\rightarrow 1$, 
 and Eq.\,(\ref{eq:highenergyB5}) reduces to, 
\bea\label{caseI}
B_5(1,a)&=&B_4(-\alpha_{AB},-\alpha_{A1})B_4(-\alpha_{23},-\alpha_{B3}+\alpha_{A1})\nonumber\\
&+&B_4(-\alpha_{AB},-\alpha_{B3})B_4(-\alpha_{12},-\alpha_{A1}+\alpha_{B3}). \nonumber \\
\eea
For $\eta\rightarrow \infty$ one can use $_1F_1(a;b;-\eta)\rightarrow (\eta)^{-a} \Gamma(b)/\Gamma(b-a)$
and obtain, 
\bea\label{caseII}
B_5(1,b)=B_4(-\alpha_{12},-\alpha_{A1})B_4(-\alpha_{23},-\alpha_{B3}).
\eea
In the following we will also take the  limit $s_{12},\,s_{23}\rightarrow \infty$  in the $B_4$ functions. 
In the double-Regge limit  specified by Eq.~(\ref{doubleRegge:condition}), four of the twelve diagrams shown in Fig.~\ref{fig:B5round} dominate. These  are the diagrams that have the same trajectory in $A1$ ($\gamma K^+$) and $B3$ ($p \bar p$) channels.  The four diagrams are related by exchanging $A\leftrightarrow1$ or $B\leftrightarrow3$ and are the 
 diagrams $B_5(1),\, B_5(7),\, B_5(12),\, B_5(9)$ in Fig.\,\ref{fig:All diagrams}. The remaining eight diagrams are 
  exponentially suppressed \cite{DoubleRegge-signature}. 

We demonstrate this suppression using the diagram $B_5(2)$ as an example. 
 Using the integral representation for the hypergeometric function, the amplitude $B_5(1)$ can be written as, 
\bea
B_5(1)&=&\frac{\Gamma(-\alpha_{12})\Gamma(-\alpha_{23})}{\Gamma(\alpha_{AB}-\alpha_{12}-\alpha_{23})}\nonumber\\
&&\times\frac{1}{2\pi i}\int_{-i\infty}^{i\infty}\frac{\Gamma(-s)\Gamma(s-\alpha_{A1})\Gamma(s-\alpha_{B3})}{\Gamma(s-\alpha_{12}-\alpha_{A1})}\nonumber\\
&&\hspace{0.5cm}\times\frac{\Gamma(s+\alpha_{AB}-\alpha_{12}-\alpha_{23})(-1)^s}{\Gamma(s-\alpha_{23}-\alpha_{B3})}ds.
\eea 
The $B_5(2)$ amplitude is obtained from $B_5(1)$  by exchanging lines $1\leftrightarrow 3$, 
\bea
B_5(2)&=&\frac{\Gamma(-\alpha_{12})\Gamma(-\alpha_{23})}{\Gamma(\alpha_{AB}-\alpha_{12}-\alpha_{23})}\nonumber\\
&&\times\frac{1}{2\pi i}\int_{-i\infty}^{i\infty}\frac{\Gamma(-s)\Gamma(s-\alpha_{A3})\Gamma(s-\alpha_{B1})}{\Gamma(s-\alpha_{23}-\alpha_{A3})}\nonumber\\
&&\hspace{0.5cm}\times\frac{\Gamma(s+\alpha_{AB}-\alpha_{12}-\alpha_{23})(-1)^s}{\Gamma(s-\alpha_{12}-\alpha_{B1})}ds,
\eea
and one finds in the double-Regge limit that it behaves as 
\bea
B_5(2) \to&& \Gamma(-\alpha_{12})\Gamma(-\alpha_{23}) \int_{-i\infty}^{i\infty}\Gamma(-s)\nonumber\\
&&\times(s-\alpha_{A3})^{\alpha_{23}}(s-\alpha_{B1})^{\alpha_{12}}(-\alpha_{AB})^{-s}ds. \nonumber \\
\eea
The amplitude thus contains the factor $e^{i\pi\alpha_{12}}e^{i\pi\alpha_{23}}$. In the 
 physical region of the reaction considered here, $\alpha_{12}$ and $\alpha_{23}$'s have positive and increasing imaginary parts which makes the amplitude $B_5(2)$ exponentially suppressed. This mechanism is analogous to the suppression of the $B_4(u,s)$ amplitude in the $s$-channel physical region at large $s$ and fixed-$t$ ({\it cf.} discussion following 
 Eq.~(\ref{bus})). 
 
\begin{figure}
\rotatebox{0}{\scalebox{0.42}[0.42]{\includegraphics{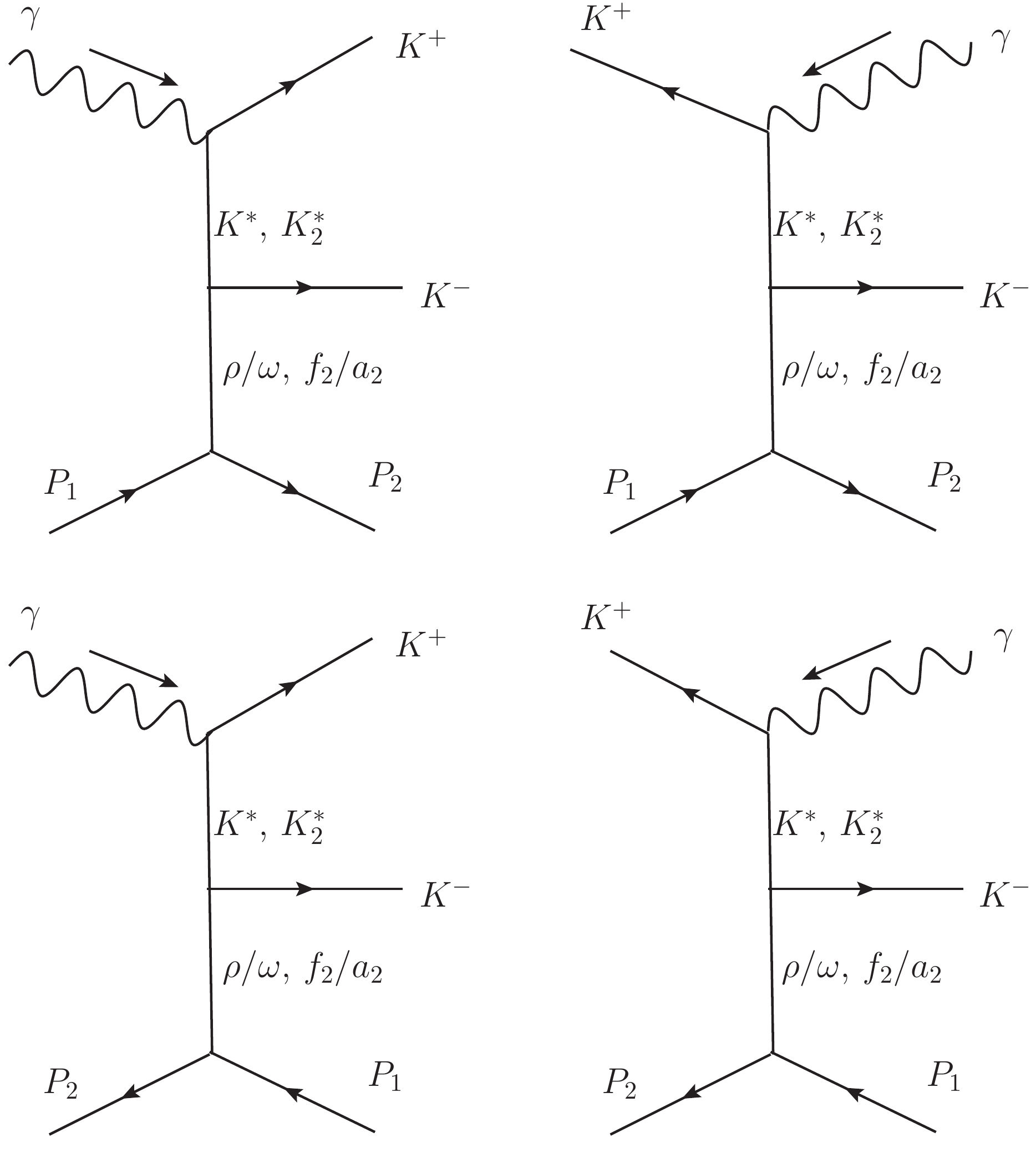}}}
\caption{The four diagrams for double-Regge limit $\gamma p\rightarrow K^+K^-p$ reaction. 
The diagrams correspond to, from top left to bottom right,  $B_5(1), B_5(7), B_5(12), B_5(9)$ in Fig.\,\ref{fig:All diagrams}, respectively. 
$K^*$ and $K_2^*$ are  the particles exchanged in the $A1$ channel while $\rho/\omega$ and $f_2/a_2$ are exchanged  
 in the  $B3$ channel.}
\label{fig:Four diagrams of double regge limit}
\end{figure}

 To obtain the DRL of the amplitudes corresponding to the diagrams $B_5(7)$ and $B_5(12)$ one  only needs to  exchange $A\leftrightarrow 1$ (upper vertex)  and $B\leftrightarrow 3$ (lower vertex), respectively. The last  diagram,  $B_5(9)$, is obtained from $B_5(1)$ by exchanging particles in both the upper and the lower vertex. The combination of the four diagrams generates the signature factors of the two Reggeons in channels $A1$ and $B3$. 
The corresponding amplitudes are given by, 
\bea
B_{5}(7) &= &B_4(-\alpha_{B1},-\alpha_{A1})B_4(-\alpha_{23},-\alpha_{B3}+\alpha_{A1})\nonumber\\
&&\times_1F_1(-\alpha_{A1};1-\alpha_{A1}+\alpha_{B3};-\eta_7)\nonumber\\
&+&B_4(-\alpha_{B1},-\alpha_{B3})B_4(-\alpha_{A2},-\alpha_{A1}+\alpha_{B3})\nonumber\\
&&\times_1F_1(-\alpha_{B3};1-\alpha_{B3}+\alpha_{A1};-\eta_7),\nonumber\\
B_{5}(12)
&=&B_4(-\alpha_{A3},-\alpha_{A1})B_4(-\alpha_{B2},-\alpha_{B3}+\alpha_{A1})\nonumber\\
&&\times_1F_1(-\alpha_{A1};1-\alpha_{A1}+\alpha_{B3};-\eta_{12})\nonumber\\
&+&B_4(-\alpha_{A3},-\alpha_{B3})B_4(-\alpha_{12},-\alpha_{A1}+\alpha_{B3})\nonumber\\
&&\times_1F_1(-\alpha_{B3};1-\alpha_{B3}+\alpha_{A1};-\eta_{12}),\nonumber\\
B_{5}(9)
&=&B_4(-\alpha_{13},-\alpha_{A1})B_4(-\alpha_{B2},-\alpha_{B3}+\alpha_{A1})\nonumber\\
&&\times_1F_1(-\alpha_{A1};1-\alpha_{A1}+\alpha_{B3};-\eta_{9})\nonumber\\
&+&B_4(-\alpha_{13},-\alpha_{B3})B_4(-\alpha_{A2},-\alpha_{A1}+\alpha_{B3})\nonumber\\
&&\times_1F_1(-\alpha_{B3};1-\alpha_{B3}+\alpha_{A1};-\eta_{9}),
\eea
where $\eta_7=\alpha_{A2}\alpha_{23}/\alpha_{B1}$, $\eta_{12}=\alpha_{12}\alpha_{B2}/\alpha_{A3}$ and $\eta_{9}=\alpha_{A2}\alpha_{B2}/\alpha_{13}$.
There are new trajectories that appear in these amplitudes. For example $\alpha_{B1}$ in $B_5(7)$ originates from $\alpha_{AB}$  in $B_5(1)$ after replacing $A$ with $1$. In principle these two trajectories have different functional dependence on the channel invariants since they represent resonances coupled to a different pair of particles. In $B_5(7)$, $\alpha_{B1}$ contains resonances in the $K^- p$ channel while $\alpha_{AB}$ in $B_5(1)$ describes resonances in the $\gamma p$ channel. As discussed in the case of the Veneziano model, to achieve resonance-Regge duality it is necessary to use a common slope for all trajectories (see discussion in Sec.~\ref{sec:DualModel}). In this case, trajectories can be related to each other using kinematical relations between channel invariants, analogous to that in Eq.~(\ref{mand}), {\it e.g.} 
\begin{equation} 
s_{AB} = s_{13}+s_{12}+s_{23}+\mbox{const.}, 
\end{equation} 
where the constant is given by sum of masses squared.
In particular one finds 
\bea\label{trajectories:relation}
\alpha_{13}&=&\alpha_{AB}-\alpha_{12}-\alpha_{23}+\mbox{const.},\nonumber\\
\alpha_{A3}&=&\alpha_{12}-\alpha_{B3}-\alpha_{AB}+\mbox{const.},\nonumber\\
\alpha_{B1}&=&\alpha_{23}-\alpha_{A1}-\alpha_{AB}+\mbox{const.},\nonumber\\
\alpha_{A2}&=&\alpha_{B3}-\alpha_{A1}-\alpha_{12}+\mbox{const.},\nonumber\\
\alpha_{B2}&=&\alpha_{A1}-\alpha_{B3}-\alpha_{23}+\mbox{const.},
\eea
which in the double-Regge limit lead to 
\bea
&&\alpha_{13}\,\,\,\sim \,\,\,\alpha_{AB}-\alpha_{12}-\alpha_{23}\,\,\,\sim \,\,\,\alpha_{AB},\nonumber\\
&&\alpha_{A3}\,\,\sim \,\,\,-\alpha_{AB}  \,\,, \,\,\,  \alpha_{B1}\,\,\,\sim \,\,\,-\alpha_{AB}  ,\nonumber\\
&&\alpha_{A2}\,\,\,\sim \,\,\,-\alpha_{12}   \,\,\,\,\,, \,\,\,    \alpha_{B2}\,\,\,\sim \,\,\,-\alpha_{23},
\eea 
 and $\eta_7 = \eta_{12}  = \eta_{9}  \to \eta$.  Combining the four surviving amplitudes in the double-Regge limit, 
\bea\label{SigI}
A_5= B_{5}(1) +\tau_{A1} B_{5}(7)  + \tau_{B3} B_{5}(12) +\tau_{A1} \tau_{B3} B_{5}(9) \nonumber \\
\eea
with $\tau_i = \pm 1$, one finds 
  \bea\label{SigII}
A_5&=&(-\alpha_{AB})^{\alpha_{A1}}(-\alpha_{23})^{\alpha_{B3}-\alpha_{A1}}\Gamma(-\alpha_{A1})\Gamma(\alpha_{A1}-\alpha_{B3}) 
 \nonumber\\
&&\times(1+\tau_{A1} e^{i\pi\alpha_{A1}}+\tau_{B3} e^{i\pi\alpha_{B3}}+\tau_{A1}\tau_{B3}e^{i\pi(\alpha_{B3}-\alpha_{A1})})\nonumber\\
& & \times V_1(t_{A1},t_{B3},\eta)  \nonumber \\
&+&(-\alpha_{AB})^{\alpha_{B3}}(-\alpha_{12})^{\alpha_{A1}-\alpha_{B3}}\Gamma(-\alpha_{B3})\Gamma(\alpha_{B3}-\alpha_{A1}) 
 \nonumber\\
&&\times(1+\tau_{A1} e^{i\pi\alpha_{A1}} +\tau_{B3} e^{i\pi\alpha_{B3}}+\tau_{A1}\tau_{B3} e^{i\pi(\alpha_{A1}-\alpha_{B3})}) \nonumber \\ & & \times V_2(t_{A1},t_{B3},\eta),
\eea
 where the functions $V_i$ represent the  Reggeon-Reggeon-particle coupling at the middle-vertex as shown in Fig.~\ref{fig:Four diagrams of double regge limit}.  The equation above has the general structure for the leading Regge pole contributions to the double-Regge limit~\cite{Brower-DeTar-Weis}, in which $V_1$ and $V_2$ are analytical functions of their variables in the kinematical domain of the double-Regge limit. In particular, the $B_5$ model used gives the following 
  prediction for the middle-vertex functions, 
\begin{eqnarray} 
V_1(t_{A1},t_{B3},\eta) &= & \,_1F_1(-\alpha_{A1};1-\alpha_{A1}+\alpha_{B3};-\eta), \nonumber \\
V_2(t_{A1},t_{B3},\eta) & = & \,_1F_1(-\alpha_{B3};1-\alpha_{B3}+\alpha_{A1};-\eta). \nonumber \\
 \label{middle} 
\end{eqnarray} 
In the double-Regge limit, the resonances-trajectories in the production channels, $\alpha_{AB}$, $\alpha_{12}$ and $\alpha_{23}$, are proportional the channel variables, 
 $\alpha_{ij}\rightarrow s_{ij}$, and 
\bea\label{SigIIImid}
A_5&=& (-s_{AB})^{\alpha_{A1}}(-s_{23})^{\alpha_{B3}-\alpha_{A1}}\Gamma(-\alpha_{A1})\Gamma(\alpha_{A1}-\alpha_{B3}) 
\nonumber\\
&&\times(1+\tau_{A1} e^{i\pi\alpha_{A1}}+\tau_{B3} e^{i\pi\alpha_{B3}}+\tau_{A1}\tau_{B3} e^{i\pi(\alpha_{B3}-\alpha_{A1})})\nonumber\\
& & \times V_1(t_{A1},t_{B3},\eta')  \nonumber \\
&+&(-s_{AB})^{\alpha_{B3}}(-s_{12})^{\alpha_{A1}-\alpha_{B3}}\Gamma(-\alpha_{B3})\Gamma(\alpha_{B3}-\alpha_{A1}) 
\nonumber\\
&&\times(1+\tau_{B3} e^{i\pi\alpha_{B3}}+\tau_{A1} e^{i\pi\alpha_{A1}}+ \tau_{A1}\tau_{B3} e^{i\pi(\alpha_{A1}-\alpha_{B3})}) \nonumber \\ & & \times V_2(t_{A1},t_{B3},\eta'),
\eea
where $\eta'= s_{12} s_{23}/s_{AB}$. We have  neglected slowly varying exponential form factors proportional to 
 $\exp( \alpha_{i} \log(\alpha'))$, where $i=A1,B3$, since for leading trajectories the slope parameter $\alpha'$ is very close to one. 

We observe, that just as in Eq.~(\ref{b4}), the above amplitude also contains ghost poles in the $A1$ and $B3$ channels for positive signatures $\tau_i = +1$ when $\alpha_{A1}$ and $\alpha_{B3}$ are the leading trajectories $\alpha(s)\sim 0.5 + s$. In the present work, we focus only on the latter case and the ghost poles have to be removed. As discussed in the previous section, these can be eliminated by shifting trajectories, $\alpha_{A1} \to \alpha_{A1} - 1$ and $\alpha_{B3} \to \alpha_{B3} - 1$. The replacement guarantees that the double-Regge amplitude has no resonances in the $A1$ and $B3$ channels in the physical region of  the $ A + B \to \bar 1 + \bar 2 + \bar 3$ reaction, where these are the exchange channels and $\alpha_{A1},\alpha_{B3} \le 0$.  Shifting the trajectories and redefining the signature accordingly we obtain
\bea\label{SigIII}
A_5&=& (-s_{AB})^{\alpha_{A1}-1}(-s_{23})^{\alpha_{B3}-\alpha_{A1}}\Gamma(1-\alpha_{A1})\Gamma(\alpha_{A1}-\alpha_{B3}) 
 \nonumber\\
&&\times(1+\tau_{A1} e^{i\pi\alpha_{A1}}+\tau_{B3} e^{i\pi\alpha_{B3}}+\tau_{A1}\tau_{B3} e^{i\pi(\alpha_{B3}-\alpha_{A1})})\nonumber\\
& & \times V_1(t_{A1},t_{B3},\eta')  \nonumber \\
&+&(-s_{AB})^{\alpha_{B3}-1}(-s_{12})^{\alpha_{A1}-\alpha_{B3}}\Gamma(1-\alpha_{B3})\Gamma(\alpha_{B3}-\alpha_{A1}) 
 \nonumber\\
&&\times(1+\tau_{B3} e^{i\pi\alpha_{B3}}+\tau_{A1} e^{i\pi\alpha_{A1}}+ \tau_{A1}\tau_{B3} e^{i\pi(\alpha_{A1}-\alpha_{B3})}) \nonumber \\ & & \times V_2(t_{A1},t_{B3},\eta').
\eea
When $\alpha_{A1}$ and $\alpha_{B3}$ are equal to even/odd positive integers, the amplitude has poles when $\tau_{A1},\tau_{B3}=+1,-1$, respectively. The amplitude in the $B3$ channel (lower Reggeon) has the first pole for   $\alpha_{B3}=1$. The pole appears in the amplitude with $\tau_{B3} = -1$ and, depending on the isospin, corresponds to the $\rho$ or $\omega$ meson exchange. 

At $\alpha_{B3} = 2$ there is a pole in the right-signature, $\tau_{B3}= +1$ amplitude. It corresponds to the lightest spin-2 tensor mesons, the $a_2$ and the $f_2$, depending on the isospin.

In the $A1$ channel, the leading pole at $\alpha_{A1}=1$ in the right-signature amplitude ($\tau_{A1} = -1$) corresponds to the exchange of the lightest strange, spin-1 meson,  the $K^*(890)$. 
The strange, tensor meson pole in the right-signature amplitude at $\alpha_{A1} = 2$ corresponds to the  $K^*_2(1430)$.  
The amplitude in the double-Regge limit is therefore associated with the exchange of the following meson combinations  
$(K^*, \rho/\omega)$, $(K^*, a_2/f_2)$ $(K^*_2, \rho/\omega)$, $(K^*_2,a_2/f_2)$,  corresponding to the amplitude with $(\tau_{A1},\tau_{B3}) = (-,-),(-,+),(+,-),(+,+)$, respectively.  
   
We note that the unnatural parity trajectory in the $A1$ channel of the $K$-meson is located  below the leading trajectory.  
 In the double-Regge limit of the $B_5$ model it is suppressed compared to the exchange of vector and tensor mesons. 
  Even though the $K$-meson exchange in the upper vertex is suppressed,  the  Pomeron exchange in the lower vertex, to which it can couple, may dominate over the leading $B3$ meson Regge exchange. Since the $B_5$ model does not include the Pomeron exchange, $K$ exchange is not considered here. 
   
Finally we also give the expressions for the DRL amplitudes in the two limiting cases discussed earlier, 
$\eta \to 0$ or $\eta \to \infty$. One finds, 
\bea\label{SigIIIcasea}
A_5&\rightarrow& (-s_{AB})^{\alpha_{A1}-1}(-s_{23})^{\alpha_{B3}-\alpha_{A1}}\Gamma(1-\alpha_{A1})\Gamma(\alpha_{A1}-\alpha_{B3}) 
 \nonumber\\
&&\times(1+\tau_{A1} e^{i\pi\alpha_{A1}}+\tau_{B3} e^{i\pi\alpha_{B3}}+\tau_{A1}\tau_{B3} e^{i\pi(\alpha_{B3}-\alpha_{A1})})\nonumber\\
&+&(-s_{AB})^{\alpha_{B3}-1}(-s_{12})^{\alpha_{A1}-\alpha_{B3}}\Gamma(1-\alpha_{B3})\Gamma(\alpha_{B3}-\alpha_{A1}) 
 \nonumber\\
&&\times(1+\tau_{A1} e^{i\pi\alpha_{A1}}+\tau_{B3} e^{i\pi\alpha_{B3}}+ \tau_{A1}\tau_{B3} e^{i\pi(\alpha_{A1}-\alpha_{B3})}),\nonumber\\
\eea
for $\eta \to 0$ and 
 \bea\label{SigIIIcaseb}
A_5&\rightarrow& (-s_{12})^{\alpha_{A1}-1}(-s_{23})^{\alpha_{B3}-1} \Gamma(1-\alpha_{A1})\Gamma(1-\alpha_{B1})
 \nonumber\\
&&\times(1+\tau_{A1} e^{i\pi\alpha_{A1}}+\tau_{B3} e^{i\pi\alpha_{B3}}+\tau_{A1}\tau_{B3} e^{i\pi(\alpha_{B3}+\alpha_{A1})}),\nonumber\\
\eea
$\eta \to \infty$, respectively.

\subsection{Spin Structure}
\label{sec:Spin Structure}

As discussed above, the dual model contains only the information about the resonance content and not about the external particles. In particular, it is agnostic about external particle spin. The amplitudes of the model should therefore 
be used in general to describe the kinematically-free scalar amplitudes appearing in a Lorentz 
decomposition of helicity amplitudes \cite{Scandron-factorliztion}. For $\gamma p \to K^+ K^- p$ 
in the expression for the helicity amplitude $M$,   
\begin{equation} 
M=\sum_{\alpha}  \bar u(p_{\bar 3},\lambda_{\bar 3}) J^\mu u(p_B,\lambda_B)  \epsilon_\mu(p_A,\lambda_A),
\end{equation}
the current $J^\mu$ is given in terms of Dirac matrices combined with the four independent particle momenta 
and multiplied by scalar functions of the invariant, Mandelstam variables. It is these scalar functions 
which can be represented by the $B_5$ amplitudes of the dual model. 

In the numerical study that follows, we test a particular model for the current operator. The model is based on the 
analysis of perturbation theory diagrams with the Reggeons replaced by lightest mass particle on the leading trajectory, \textit{i.e.} 
the vector mesons. This is most accurate for the $(\tau_{A1},\tau_{B3})=(-,-)$ amplitude, \textit{c.f.} discussion above, 
while the other three combinations should include at least one tensor structure associated with the exchange of a tensor meson. 
We note that exchange of higher-spin states does not require further modification of the spin-tensors, since as far as the spin-structure is concerned, 
the only difference between \textit{i.e.} spin-$3$ and spin-$1$ meson exchange is an analytical function of the   
channel variables. 
The dependence on these variables is already fixed by the dual Regge limit. 
In the following, we make a simplifying approximation and use the same spin structure for all four signature combinations.  

The (upper) vertex representing a coupling of an external (vector) photon to a (pseudoscalar) kaon, via exchange of a vector meson in the $A1$ channel is given by 
\begin{equation} 
\label{vectorV}
V^u(\lambda_{A1})   = \epsilon_{\mu\nu\alpha\beta} \epsilon_\mu(p_A,\lambda_A) \epsilon^\nu(p_{A1},\lambda_{A1}) p_{1}^\alpha p_A^\beta.
\end{equation}
Here $p_{A1} = p_{A} + p_1$ and $\lambda_A$ and $\lambda_{A1}$ are the helicities of the photon and the exchanged vector meson, respectively. 
Similarly, the bottom vertex represents the coupling of a vector meson in the $B3$ channel to the two nucleons is given either by 
\begin{equation} 
V^l_I(\lambda_{B3})  = \epsilon^\mu(p_{B3},\lambda_{B3}) \bar u(p_{\bar 3},\lambda_{\bar 3})\gamma_\mu u(p_B,\lambda_B)
\end{equation} 
or by 
\begin{equation} 
V^l_{II}(\lambda_{B3})  = \epsilon^\mu(p_{B3},\lambda_{B3}) \bar u(p_{\bar 3},\lambda_{\bar 3}) i\sigma_{\mu\nu}p_{B3}^\nu u(p_B,\lambda_B) \label{vectorl},
\end{equation} 
with $p_{B3} = p_A + p_B$ and $\lambda_{B3}$ representing momentum and helicity of the exchanged vector meson in the $B3$ channel. 
 The $V^l_{I/II}$ vertex represents dominantly the helicity-flip/non-flip amplitude in the $s$-channel, respectively. 
 The $\rho$ meson exchange, for example,  is expected to be dominantly helicity flip and corresponds to the vertex $V^l_{II}$. 
 In the following, we will use $V^l_{II}$ for the bottom vertex for the four Reggeon combinations discussed in the preceding 
 section. 

In Eq.~(\ref{middle}), the vertex functions $V_i$  describes the middle-vertex where the $A1$ and $B3$ 
exchanges couple to the external particle. At the exchanged particle poles, $V_i=1$, 
the amplitude has to be multiplied by the appropriate Clebsch-Gordan coefficient representing, in our case, 
the coupling of the two exchanged vectors to the pseudoscalar kaon. 
 This coupling is given by 
\begin{equation} 
V^m(\lambda_{A1},\lambda_{B3}) = \varepsilon_{\mu\nu\alpha\beta}  \epsilon^\mu(p_{A1},\lambda_{A1}) \epsilon^\nu(p_{B3},\lambda_{B3})  p_{A1}^\alpha p_2^\beta.  \label{vectorm} 
\end{equation} 

The final amplitude  $M$ is obtained by multiplying $A_5$ by the product of the three vertices, Eqs.~(\ref{vectorV},~\ref{vectorl},~\ref{vectorm}), and summing over helicities of the intermediate vector exchanges,  
\begin{equation} \label{finalamp}
M = A_5 \sum_{\lambda_{A1},\lambda_{B3}} V^u(\lambda_{A1}) V^m(\lambda_{A1},\lambda_{B3}) V^l_{II}(\lambda_{B3}).
\end{equation} 
When the spin is averaged, the spin-amplitudes produce an intensity function that is a regular function of the Mandelstam variables.  The key feature of the double-Regge limit resides in the predicted dependence of the sub-channel energies, $s_{12},s_{23}$ and $s_{AB}$, which,  far away from the resonance region is quite smooth. 
In the following we analyze the limit by studying the Dalitz plot distributions guided by the selection criteria proposed in ~\cite{VanHove-shortpaper,VanHove-longpaper} based on the analysis of the longitudinal component of the momentum, also  called  longitudinal or Van Hove plots.

\section{Data Simulation}\label{sec:DataSimulation}

\begin{figure}
\rotatebox{0}{\scalebox{0.42}[0.42]{\includegraphics{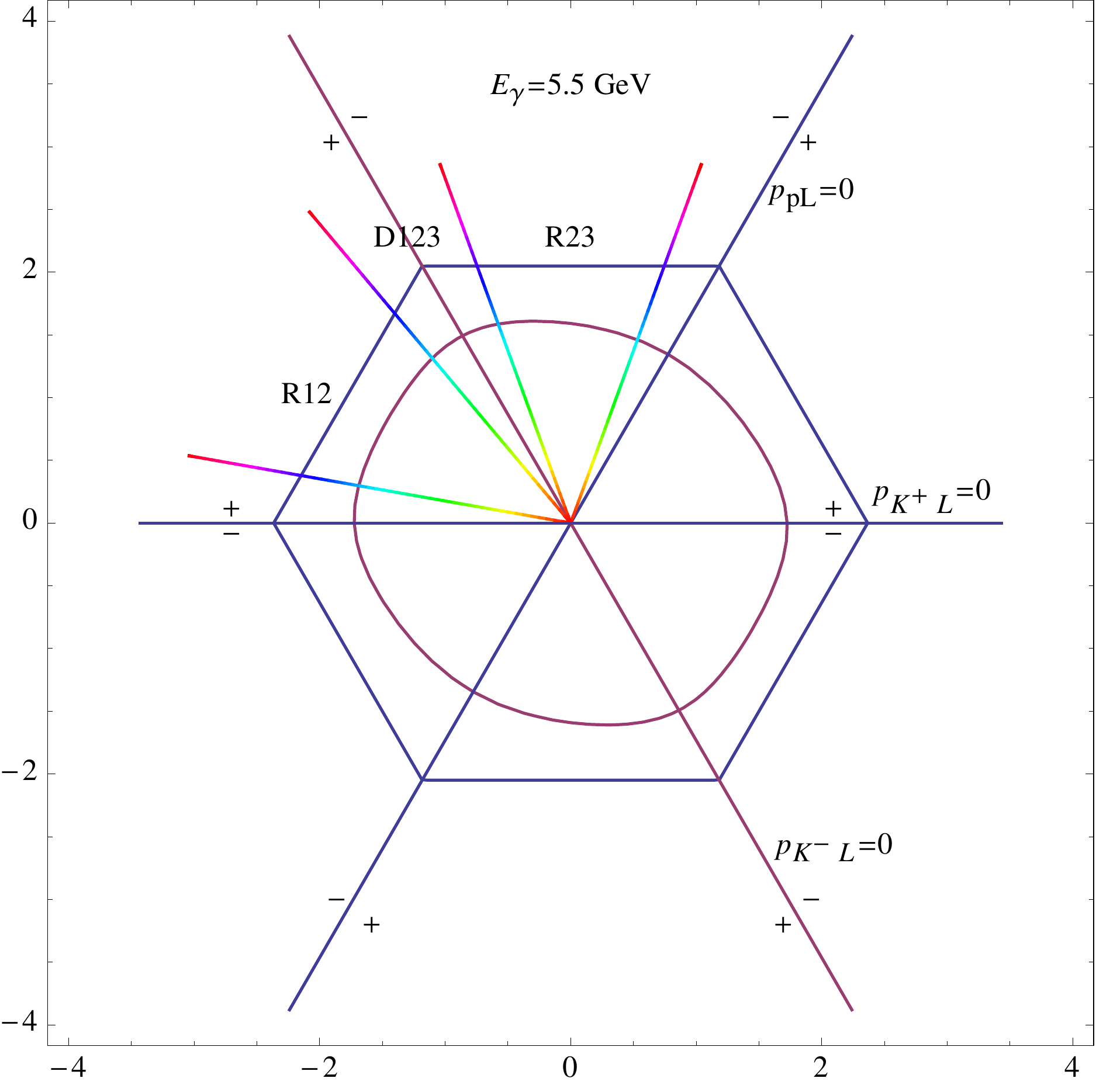}}}
\caption{The boundaries of the Van Hove plot for the $\gamma p\rightarrow K^+K^-p$ reaction. 
The chosen photon energy is $5.5$ GeV.
The \textquoteleft$+/-$\textquoteright sign stands for either the parallel (+) orthe 
anti-parallel ($-$) direction of the outgoing $K^+$, $K^-$, or $p$ momenta compared to the photon momentum in the c.o.m. frame. }
\label{fig:VanHove}
\end{figure}

In this section, we use the amplitude obtained above, Eq. (\ref{finalamp}), to simulate the $\gamma p\rightarrow K^+K^-p$ reaction.  
We set the photon energy to $E_\gamma = 5.5$ GeV ($\sqrt{s}=3.5$ GeV)  in the lab-frame (target rest frame)  which corresponds to the  highest photon energy of the $K^+ K^-$  data collected by the CLAS collaboration at JLab~\cite{CLAS2014,JlabE04005}. 
Once the data are analyzed they can be compared with the simulations presented below.  
We use a standard Monte Carlo (MC)  method to generate the events according the the amplitudes discussed above.

For the leading trajectories, as discussed in Sec.~\ref{B5amp}, we use meson trajectories from \cite{all-trajectories}
\begin{equation}
\begin{split}
\alpha_{A1}(t)&=\alpha_{K^*} = 0.318+0.839t,\\
\alpha_{B3}(t)&=\alpha_{\rho}= 0.456+0.887t.
\end{split}\label{eq:trajectories}
\end{equation}
The numerical computation of the $A_5$ amplitude 
must be done carefully when evaluating the Regge-Regge-particle vertices given by Eq.~(\ref{finalamp}). 
Each term separately is singular when 
$\alpha_{A1}$ and $\alpha_{B3}$ are not integers but $\alpha_{A1}-\alpha_{B3}$ 
is an even integer, although the full amplitude is finite at those points.
To avoid singularities in the $V_i$ we add a small imaginary part to the $\alpha_{A1}$ and $\alpha_{B3}$ Regge trajectories,
shifting the location of the poles outside the real axis where the amplitude is evaluated. 

We study four amplitudes defined by the four possible combinations of the signature factors, which as discussed in Sec.~\ref{B5amp}, 
correspond to the following cases:
\begin{itemize}
\item  [I:] $\tau_{A1}=-1$, $\tau_{B3}=+1$, for ($K^*,a_2/f_2$);
\item  [II:] $\tau_{A1}=-1$, $\tau_{B3}=-1$, for ($K^*,\rho/\omega$);
\item  [III:] $\tau_{A1}=+1$, $\tau_{B3}=+1$, for ($K_2^*,a_2/f_2$);
\item  [IV:] $\tau_{A1}=+1$, $\tau_{B3} =-1$, for ($K_2^*,\rho/\omega$).
\end{itemize}
 In this analysis we do not distinguish between different isospins, but we do
 study the spin structure described in Sec.~\ref{sec:Spin Structure}.

\subsection{Data Selection and the Van Hove Plot}
\label{sec:Van Hove Plot}
At fixed $s_{AB}$, we integrate over $t_{A1}$ and $t_{B3}$, 
the cross-section then becomes a function of $s_{12}$ and $s_{23}$ only and 
can be represented in a Dalitz plot.  The double-Regge limit corresponds to low values of the momentum transfer variables, $t_{A1}$ and $t_{B3}$. To isolate the corresponding DRL in the Dalitz variables it is best to employ the procedure developed by Van Hove ~\cite{VanHove-shortpaper,VanHove-longpaper}.

\begin{figure}
\rotatebox{0}{\scalebox{0.42}[0.42]{\includegraphics{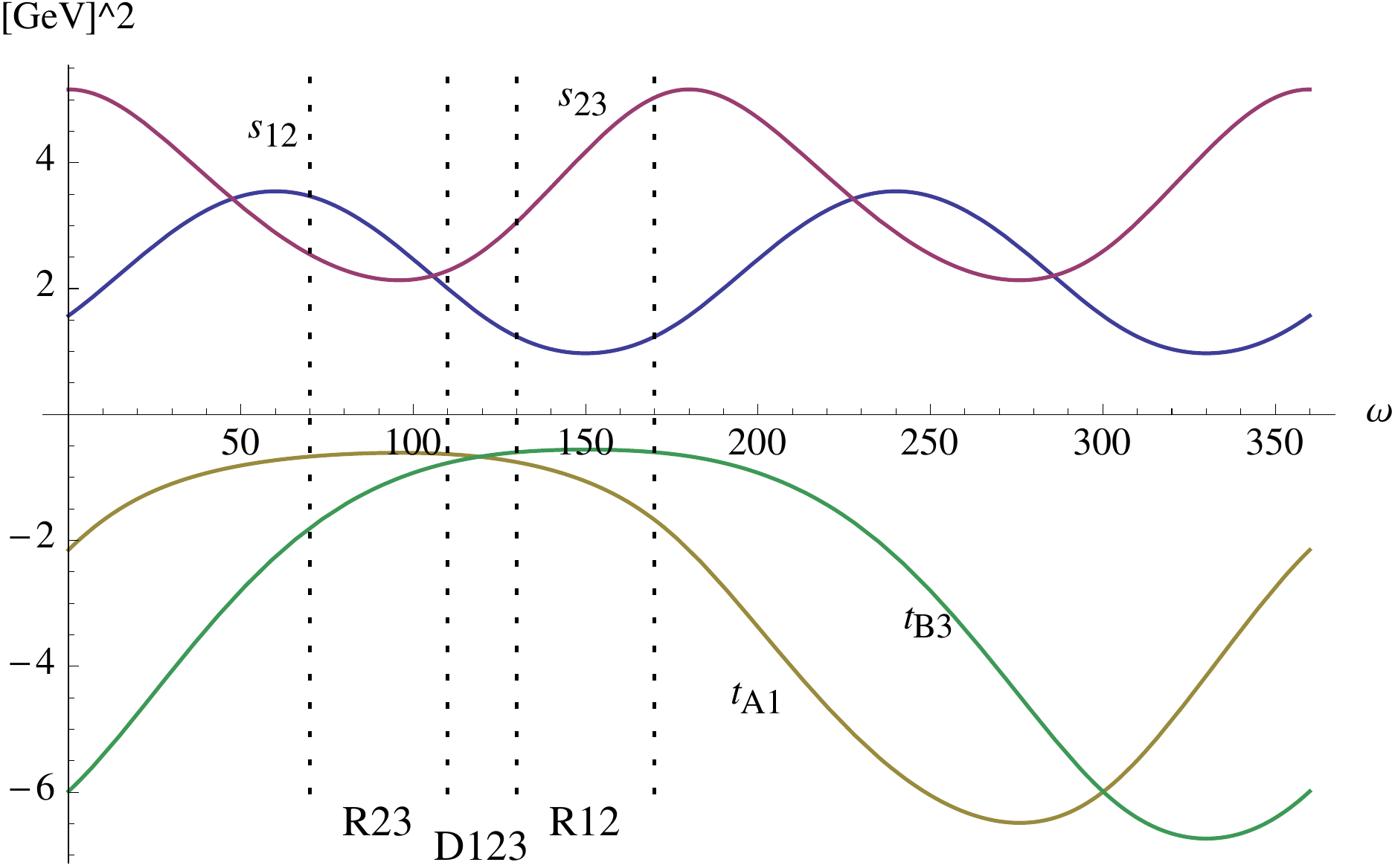}}}
\caption{The variables $s_{12}$ (blue), $s_{23}$ (purple), $t_{A1}$  (yellow), and
$ t_{B3}$ (green) as
functions of the polar angle $\omega$ for $\gamma p\rightarrow K^+K^-p$ 
at photon energy 5.5 GeV ($s_{AB}=11.2 \mbox{ GeV}^2$). 
The transverse momenta are set to $|p_{K^+T}|=|p_{K^-T}|=|p_{PT}|=0.6$ GeV. }
\label{fig:VanHoveVaribales}
\end{figure}

For a $2-$to$-3$ process, the Van Hove plot is a two-dimensional plot of the longitudinal  momenta of the three produced particles. In the center of mass (c.o.m.) frame, the incident photon defines the positive-$z$ axis and the longitudinal components of the outgoing particles are defined by the projection of the
momenta onto the $z$ axis. Longitudinal momentum conservation mandates that only two out of the three particle's longitudinal momenta are independent.  Furthermore, energy and transverse-momentum conservation require that events are distributed inside a bounded region of the two dimensional space defined by the independent longitudinal momenta. 
 The longitudinal momenta of the outgoing particles are parameterized using polar coordinates 
 with  radius $q=(p_{K^+L}^2+p_{K^-L}^2+p_{PL}^2)^{\frac{1}{2}}$ and a  polar angle
   $\omega$ defined as: 
\begin{equation}
\begin{split}
p_{K^+L}&=\sqrt{\frac{2}{3}} q \sin \omega \,,\\
p_{K^-L}&=\sqrt{\frac{2}{3}} q \sin\left( \frac{2}{3}\pi+\omega\right)\,,\\
p_{PL}&=\sqrt{\frac{2}{3}} q \sin\left(\frac{4}{3}\pi+\omega\right)\, .
\end{split}
\label{logitudinal_momentum}
\end{equation}
With the lines corresponding to $p_{K^+L} = 0$, $p_{K^-L}=0$, and $p_{PL}=0$ drawn at $60^0$ angle on a 2-dimensional plot, {\it aka} longitudinal plot, {\it cf.} Fig.~\ref{fig:VanHove}. Each point satisfies longitudinal momentum conservation.  
 In the limiting case where particle masses  and transverse momenta are ignored, the boundary of the longitudinal plot 
for  $\gamma p\rightarrow K^+K^-p$  corresponds to the hexagon in Fig.~\ref{fig:VanHove}.  
Otherwise it is given by the smooth curve shown by the inner elliptical-line defined by vanishing transverse momenta.

\begin{figure}
\subfigure[\ Generated phase space events.]{
\rotatebox{0}{\scalebox{0.42}[0.42]{\includegraphics{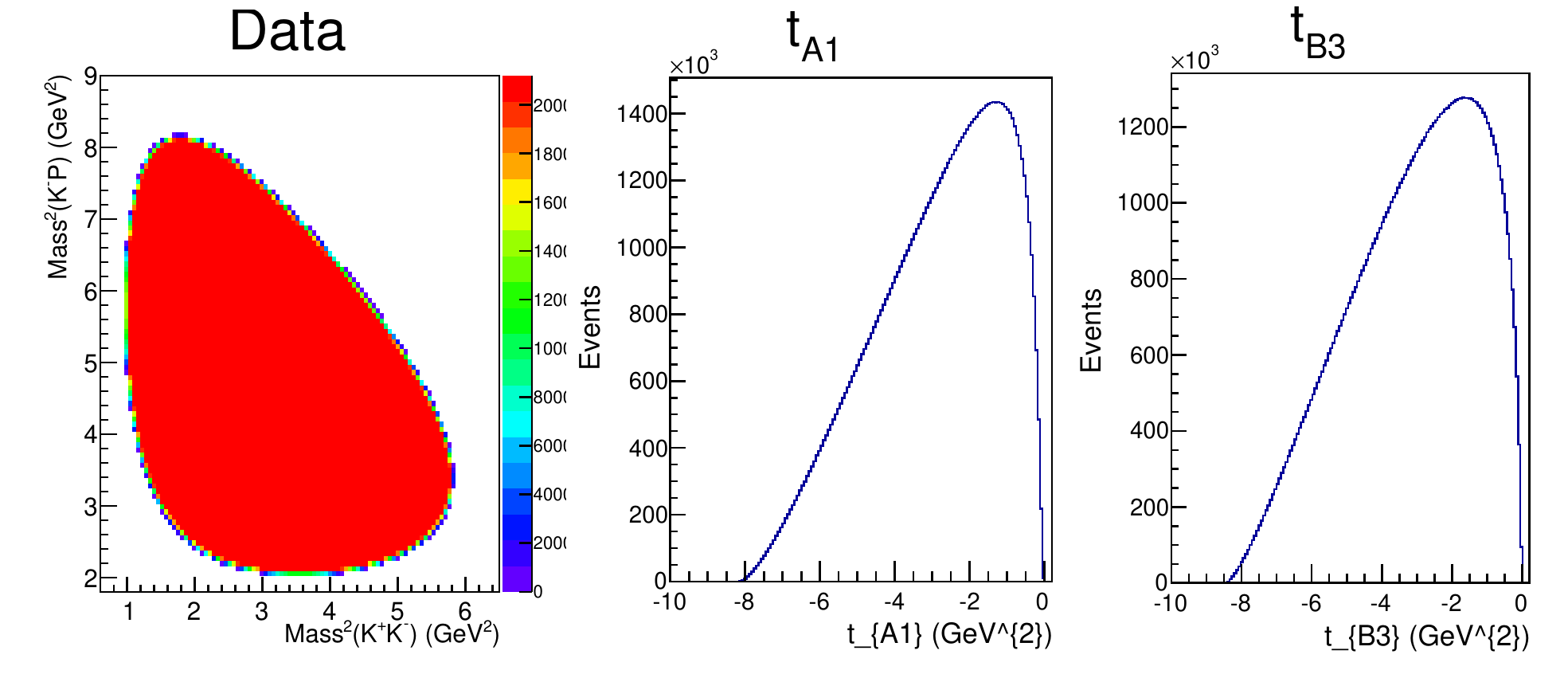}}} \label{Fig:phasespace}} 
\subfigure[\ Phase space events after transverse momenta cut.]{
\rotatebox{0}{\scalebox{0.42}[0.42]{\includegraphics{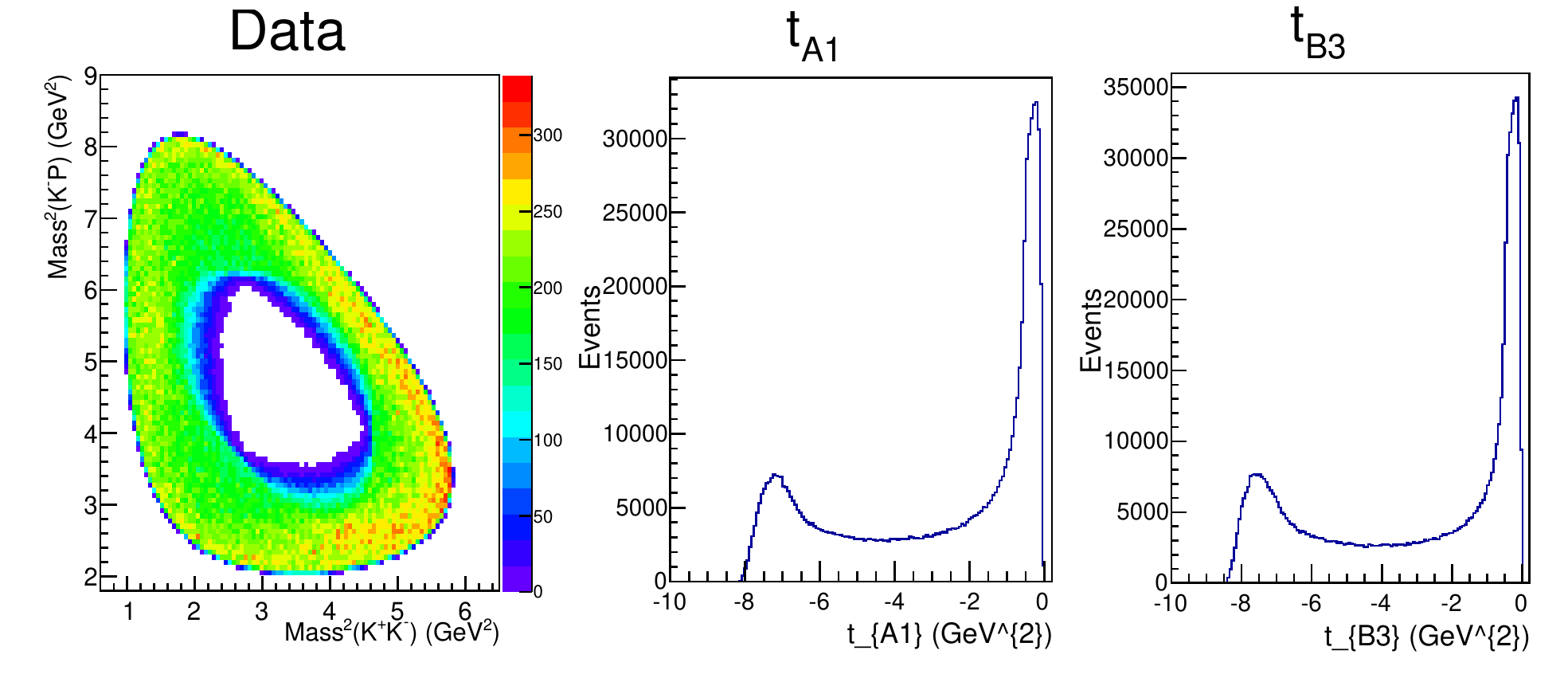}}} \label{Fig:phasespacept} } 
\subfigure[\ Phase space events for the double-Regge limit.]{
\rotatebox{0}{\scalebox{0.42}[0.42]{\includegraphics{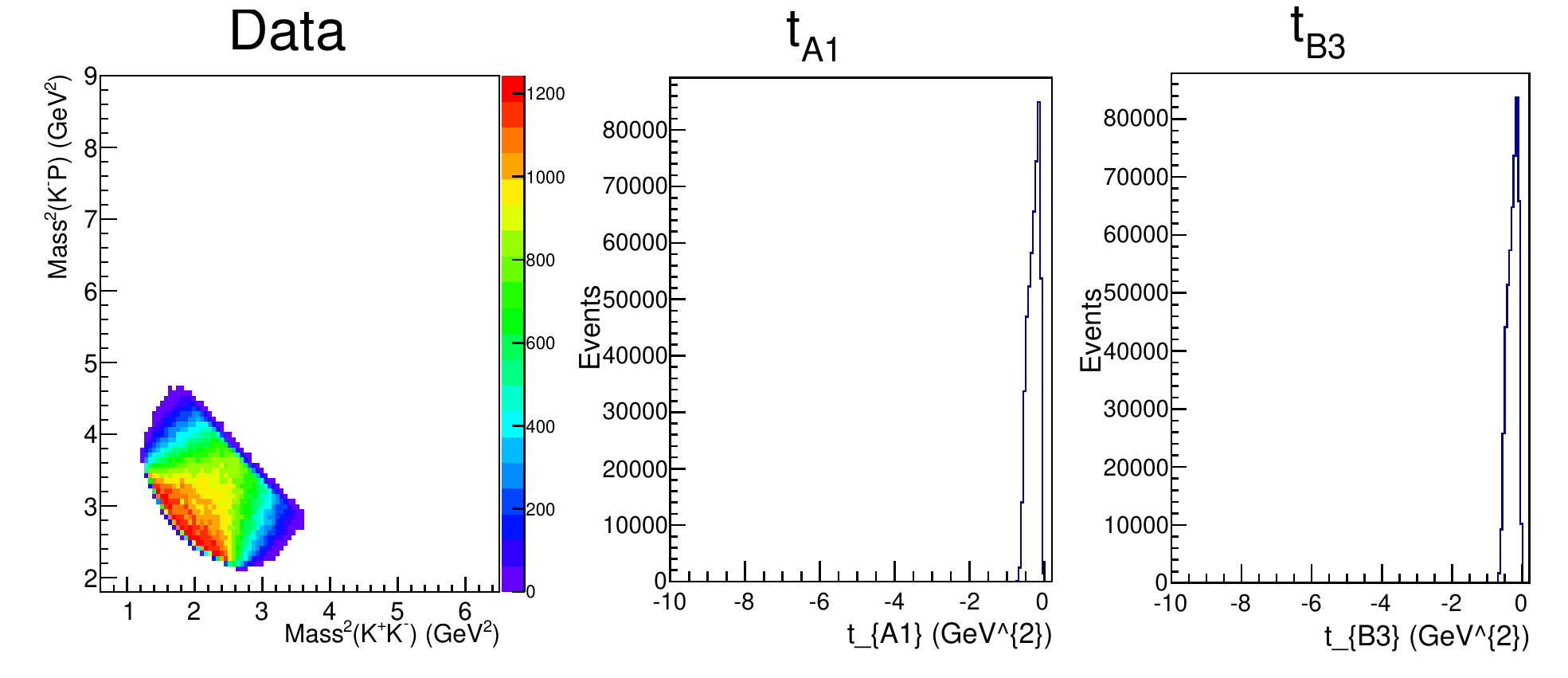}}}  \label{Fig:psptVH}}
\caption{(a) Generated phase space events. The three plots depict (from left to right)
the Dalitz plot and the $t_{A1}$ and $t_{B3}$ distributions;
(b) Phase space events after the transverse momenta $p_{TK^+}$, $p_{TK^-}$, and $p_{PT}$ have been constrained within the $[0, 0.6]\: \mbox{GeV}$ region;
(c) Phase space events for the double-Regge limit after performing the transverse momenta cuts and Van Hove selection described in Sec. \ref{sec:Van Hove Plot}.}
\end{figure}

With the parametrization given in Eq.~(\ref{logitudinal_momentum}),  the Dalitz variables,  $s_{12}$ and $s_{23}$, and the two momentum transfer, $t_{A1}$ and $t_{B3}$, become functions of $\omega$ and the transverse momenta. 
For example, given a photon beam energy of $5.5 \mbox{ GeV}$, which corresponds to $s_{AB}=11.2 \mbox{ GeV}^2$ 
and fixing  $|p_{K^+T}|=|p_{K^-T}|=|p_{PT}|=0.6 \mbox{ GeV}$, the $\omega$ dependence 
of the Dalitz and momentum transfer variables are
shown in Fig.~\ref{fig:VanHoveVaribales}.

The longitudinal momentum $p_{K^+L}$ is positive for $\omega \in [0^\circ,180^\circ]$ and  $p_{PL}$ is negative for $\omega \in  [60^\circ,240^\circ]$, hence, the events concentrate in the upper left and center sectors of the plot in Fig. \ref{fig:VanHove}. The overlap of these two regions  $\omega \in  [60^\circ,240^\circ]$  corresponds to low momentum transfers in $A1$ and $B3$ channels.

 The Dalitz variables $s_{12}$ and $s_{23}$ are periodic with $180^\circ$ period and their minima in the region that overlaps with the region of small momentum transfer are $\omega \in [ 110^\circ,  170^\circ]$ for $s_{12}$ and $\omega \in [ 70^\circ,  110^\circ]$ for  $s_{23}$. These define the single-Regge limits that correspond to large $s_{AB}$ and small,  either, $s_{12}$  or $s_{23}$. These are labeled as the R12 and R23 wedges in Fig.~\ref{fig:VanHove}.

The region were both $s_{12}$ and $s_{23}$ are largest that overlaps with the region of small momentum transfers 
 corresponds to $\omega \in [ 110^\circ, 130^\circ]$. This is the region which is closest to the kinematic domain 
  of the double-Regge limit and it is marked by the wedge labeled  D123 in  Fig.~\ref{fig:VanHove}.
In this region, in the center of mass frame, the $K^+$ and  the recoiling proton have large momentum components in the 
 $+z$ and $-z$ direction, respectively while the 3-momentum components of the $K^-$ are small.

To generate events in the double-Regge region we use the following procedure. 
First, we generate a large sample $O(10^8)$ events uniformly distributed in the three particle phase space. 
In Fig.\,\ref{Fig:phasespace} we show the  generated Dalitz plot and the distribution of the momentum transfer 
  $t_{A1}$ and $t_{B3}$.
Next, we limit the momentum transfers by constraining the transverse momenta $|p_{K^+T}|$, $|p_{K^-T}|$, and $|p_{PT}|$ to the $[0,0.6] \, \mbox{GeV}$ range.
The center of the longitudinal plot in Fig.~\ref{fig:VanHove} corresponds to vanishing longitudinal momenta $q=(p_{K^+L}^2+p_{K^-L}^2+p_{PL}^2)^{\frac{1}{2}}=0$ and maximal value  of the transverse momenta. 
Because of the cut off on the value of transverse momenta, a hole appears in the longitudinal plot 
  which also  shows up in the Dalitz plot in Fig. \ref{Fig:phasespacept}.  The middle and right panel in Fig. \ref{Fig:phasespacept} shows that, despite of the cuts in transverse momenta, there are still contributions from 
   large $|t_{A1}|$ and $|t_{B3}|$ that need to be removed. As discussed above, the contribution from the large momentum transfers is eliminated by restricting $\omega$ to the D123 region of 
    $\omega\in[110^\circ,130^\circ]$.  After the cut on $\omega$ events with momentum transfers, $|t_{A1}|$ and $|t_{B3}|$, larger than $1.5\ \mbox{GeV}^2$ are removed as shown in 
     in  Fig. \ref{Fig:psptVH}. The final sample is reduced to approximately $5 \times 10^6$ events. 
  The Dalitz plot distribution is shown in the left panel in  Fig.~\ref{Fig:psptVH} and it agrees with that of \cite{DoubleRegge-kinematic}.

\begin{figure}
\rotatebox{0}{\scalebox{0.42}[0.42]{\includegraphics{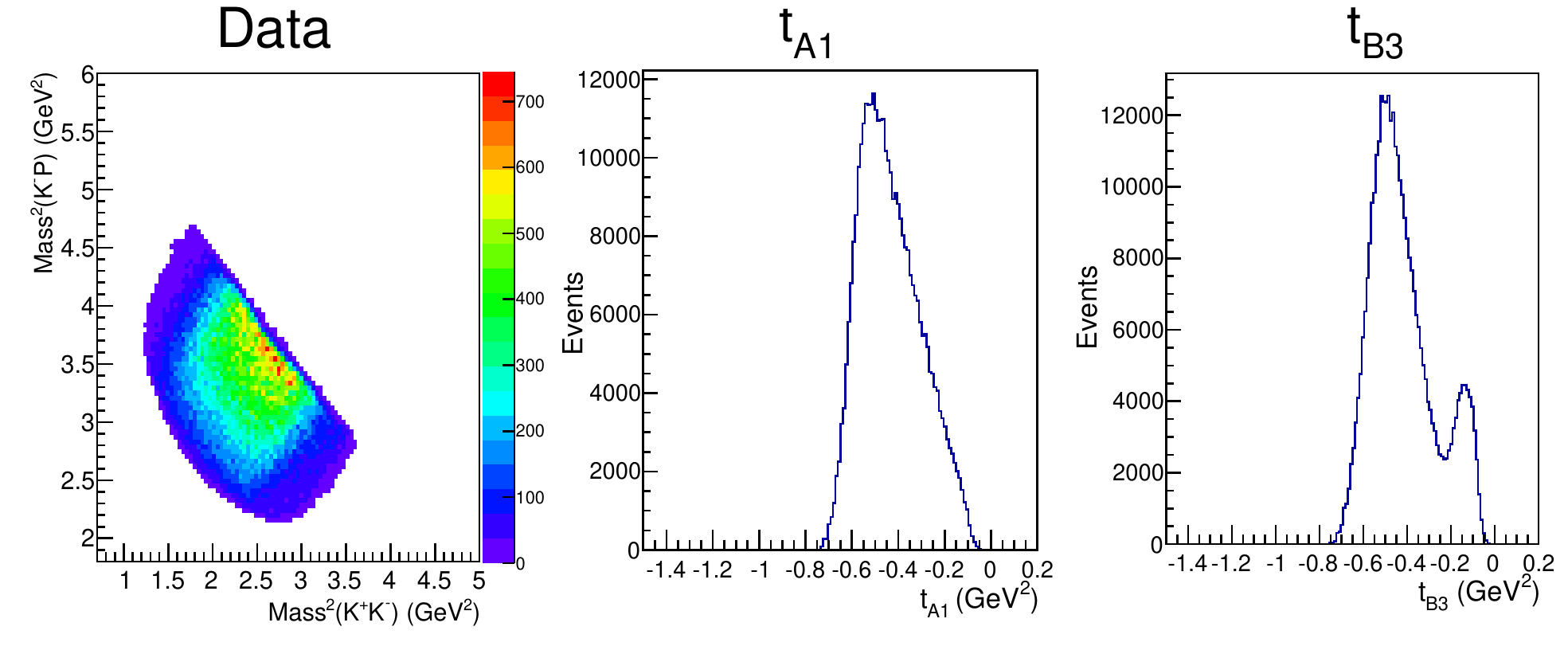}}}
\caption{Event distribution due to the spin structure after  performing the transverse momenta cuts and 
Van Hove selection. }
\label{Fig:spinstructureVH}
\end{figure}

The final amplitude depends on the spin structure of the external particles as explained in Sec. \ref{sec:Spin Structure}. 
In order to study the spin structure, in Fig. \ref{Fig:spinstructureVH} we plot the amplitude $M$ in Eq. (\ref{finalamp}) with $A_5=1$ for the selected events. 
Comparing the Dalitz plots in Fig. \ref{Fig:psptVH} and Fig. \ref{Fig:spinstructureVH} we notice that  
event concentration shifts from the bottom-left of the Dalitz plot to the center. 
Fig. \ref{Fig:spinstructureVH} also shows that the kinematical factor suppressed the events in the forward direction 
 due to the spin-flip nature of the bottom vertex.

\begin{figure}
\rotatebox{0}{\scalebox{0.42}[0.42]{\includegraphics{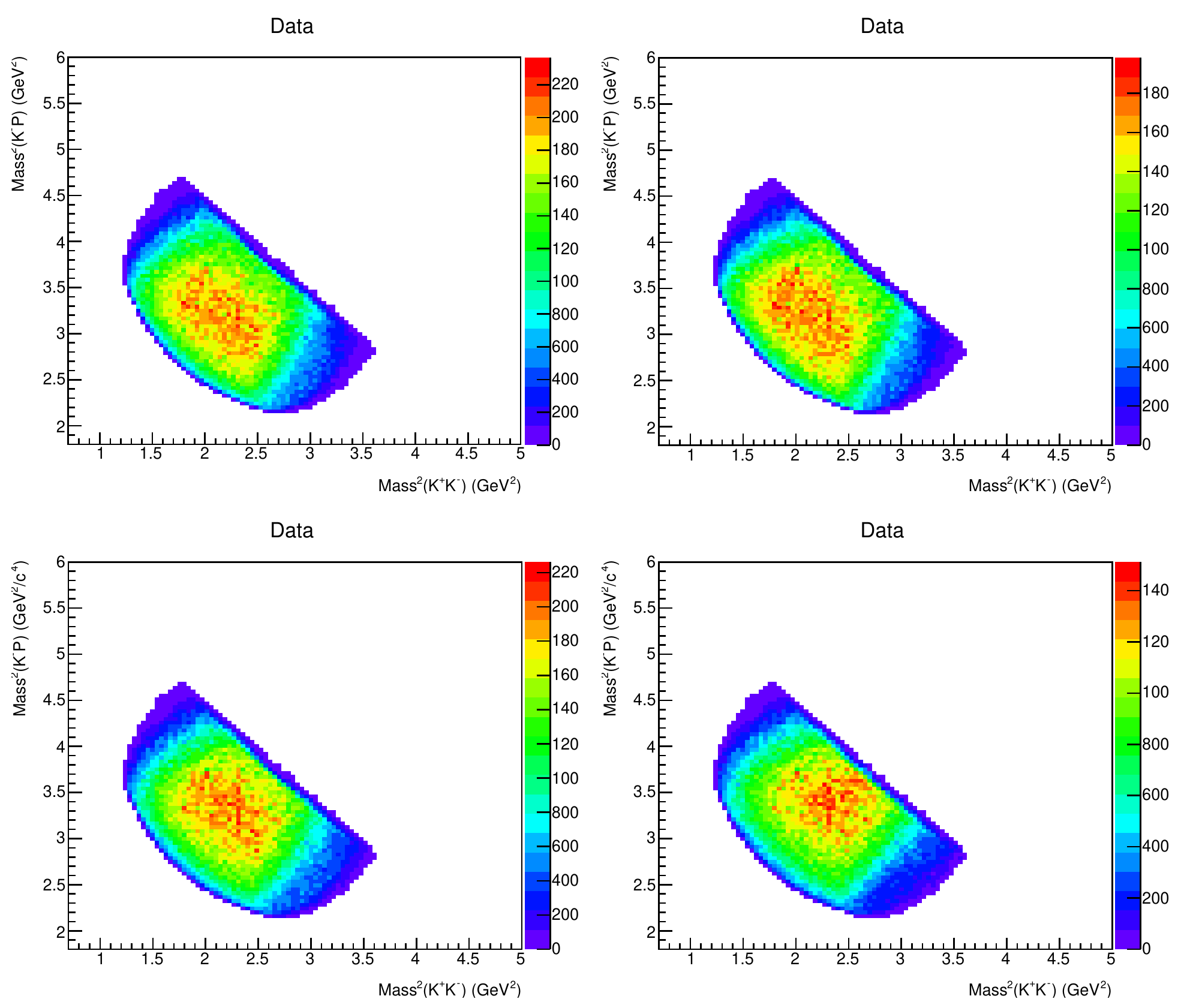}}}
\caption{Generated MC data for the double-Regge amplitude in Eq. (\ref{finalamp})  for the four cases described in Sec. \ref{sec:DataSimulation}: case I (top left), case II (top right), case III (bottom left), case IV (bottom) right.}
\label{Fig:gen}
\end{figure}

Once we have chosen the kinematical region carefully and we understand the impact of the spin structure in the amplitude, 
we can simulate the full amplitude $M$ in  Eq. (\ref{finalamp}) for the four cases described in Sec. \ref{sec:DataSimulation}.
The four cases have similar, but distinguishable, characteristics that are apparent in Dalitz plots shown in Fig.\ref{Fig:gen}. 
For instance, in all the cases, the events are concentrated in the middle of the selected kinematical region. 
All four amplitudes share the same spin structure and the same dependence on $s_{12}$ and $s_{23}$ but have different combinations of exchanged Reggeons. 
Because of the small $|t_{A1}|$ and $|t_{B3}|$  the signature factors become approximately constant and therefore 
 do not introduce significant differences among the four amplitudes, as shown in Fig.~\ref{Fig:gen}.

\begin{figure}
\rotatebox{0}{\scalebox{0.42}[0.42]{\includegraphics{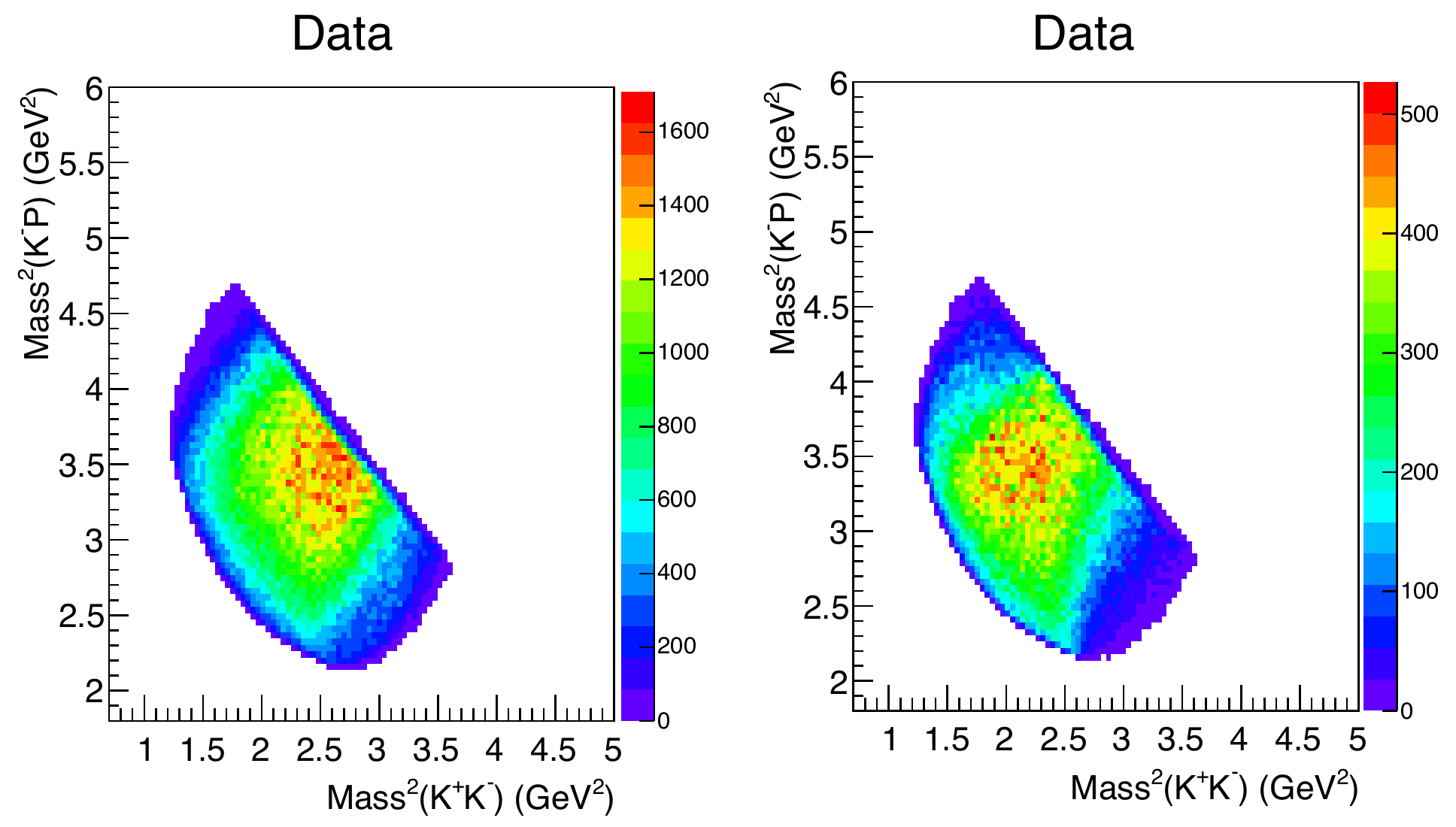}}}
\caption{Double-Regge limit amplitudes for case I under the limits: $\eta\rightarrow 0$ (left), $\eta\rightarrow \infty$ (right). }
\label{Fig:acceta}
\end{figure}

Finally we discuss the results for the limits,  $\eta\rightarrow 0$ and $\eta\rightarrow \infty$, corresponding to the amplitudes in 
 Eqs. (\ref{SigIIIcasea}) and (\ref{SigIIIcaseb}). 

Specifically we take $\tau_{A1}=-1$ and $\tau_{B3}=+1$, corresponding to $K^*$ and $a_2/f_2$, resulting in  Fig. \ref{Fig:acceta}. As expected the limit $\eta\rightarrow\infty$ favors events which maximize the channel invariants in the middle of Dalitz plot.

\section{Conclusions and Outlook}\label{sec:conclusion}
 Experiments have been performed and are planned for the future in which multiparticle final states are produced with a view to identifying new resonances, as well as confirming the properties of previously found states. Having only nucleon targets, these final states inevitably involve both baryon and meson resonances.  A  common analysis procedure adopted is to regard the mesons as an unwanted background to the study of baryon resonances, and the $N^*$'s (or here the $\Lambda$'s and $\Sigma$'s)  as an annoying background to the study of meson resonances. Kinematic cuts are made to eliminate one or the other, and so enhance the baryon or meson signal as required. Of course, this has the deficiency of throwing away not just a `background', but part of the signal too, as well as the essence of the production process. Dalitz plot analyses together 
  with analysis of momentum transfer distribution highlight how the overlap of kinematic regimes contains valuable information that can help to elucidate the signals under study. 
  
  In this article we have studied the double-Regge exchange limit for the $\gamma p\rightarrow K^+K^-p$ reaction employing a generalized Veneziano model ($B_5$ model).
The equations necessary to describe the double-Regge limit are obtained by taking the high-energy limit, $s_{AB}, s_{12}, s_{23}\rightarrow \infty$, as well as restricting the momentum transfers. 
When the 2-to-3 amplitude is saturated by the two Regge poles, the dependence on the sub-channel energies is fixed by the Regge pole trajectories and can be tested by studying the Dalitz plot distributions as a function of the (small) momentum transfers.    
We have shown that suitable event candidates for the double-Regge exchange in the high-energy limit 
can be selected by means of longitudinal momentum distribution which provide the adequate cuts in the sub-channel invariants and momentum transfers. 
The importance of the spin structure in the amplitude has been investigated and it was found that it may have impact on the density distribution of the Dalitz plot.  
We have identified and simulated four leading two Regge pole exchanges 
($K^*,a_2/f_2$, $K^*,\rho/\omega$, $K_2^*,a_2/f_2$, and $K_2^*,\rho/\omega$).
Because of the small range of momentum transfer the signature factors become approximately constant and the leading trajectories are approximately exchange degenerate. Consequently, we find little sensitivity to exchange dynamics.

 The double-Regge limit operates in the kinematic regime where the two sub-channel invariants are large and outside the resonance region in each channel. Analyticity implies that amplitude in the resonance region, in both meson and baryon channels should connect smoothly to the amplitude in double-Regge kinematics. 
 This constrain can be formalized using finite energy sum rules~\cite{Hoyer:1973hf}. 
Having a  common framework to analyze both baryonic and mesonic signals is key to making combined  multiparticle analyses tractable. The isolation of the double-Regge limit is a step in that direction by providing  a realistic and accurate modeling that feeds into both the dynamics of mesons and baryons.  The single Regge limit and the above mentioned analytical continuation  for  $\gamma p\rightarrow K^+K^-p$ will be presented in future works.

\begin{acknowledgments} 

We thank V. Mokeev and C. Salgado for useful comments.
We also thank W. F. Perger 
for providing the hypergeometric function code. 
Meng Shi stay at Jefferson Lab was supported by the CSC scholarship of  the Chinese government.
This material is based upon work supported by the U.S. Department of Energy, Office of Science, Office of Nuclear Physics under contract DE-AC05-06OR23177.
This work was also supported in part by the U.S. Department of Energy under Grant No. DE-FG0287ER40365, National Science Foundation under Grant PHY-1415459.
\end{acknowledgments} 

\appendix

\section{Useful Relations} \label{sec:appendix}
Throughout this manuscript we make extensive use of the hypergeometric function defined by 
\begin{widetext}
\begin{equation}
_pF_q(x_1,\cdots,x_p;y_1,\cdots,y_q;z) = 
\sum^{\infty}_{k}\frac{\Gamma(x_1+k)\cdots\Gamma(x_p+k)\Gamma(y_1)\cdots\Gamma(y_q)}{\Gamma(x_1)\cdots\Gamma(x_p)\Gamma(y_1+k)\cdots\Gamma(y_q+k)} \frac{z^k}{k!},
\end{equation}
and its properties.
If we write $_pF_q(x_1,\cdots,x_p;y_1,\cdots,y_q)= \, _pF_q(x_1,\cdots,x_p;y_1,\cdots,y_q;1)$, for $z=1$, the series converges if $\text{Re}(\sum y_q - \sum x_p) > 0$. 
Outside of its domain of  convergence, the hypergeometric function is defined by analytical continuation, which can be performed aided by the following relations
\begin{equation}
\begin{split}
_3F_2(a,b,c;d,e)&=
\frac{\Gamma(1-a)\Gamma(d)\Gamma(e)\Gamma(c-b)}{\Gamma(e-b)\Gamma(d-b)\Gamma(1+b-a)\Gamma(c)}\times\, _3F_2(b,1+b-d,1+b-e;1+b-c,1+b-a)\\ 
&+\frac{\Gamma(1-a)\Gamma(d)\Gamma(e)\Gamma(b-c)}{\Gamma(e-c)\Gamma(d-c)\Gamma(1+c-a)\Gamma(b)} \times\, _3F_2(c,1+c-e,1+c-d;1+c-b,1+c-a),
\end{split}\label{eq:HYPrelation1}
\end{equation}

\begin{equation}
\frac{_3F_2(a,b,c;d,e)}{\Gamma(s)\Gamma(d)\Gamma(e)}=\frac{_3F_2(d-c,d-b,a;s+a,d)}{\Gamma(e-a)\Gamma(s+a)\Gamma(d)}
=\frac{_3F_2(s,d-a,e-a;s+b,s+c)}{\Gamma(a)\Gamma(s+b)\Gamma(s+c)}, \label{eq:HYPrelation2}
\end{equation}
\end{widetext}
where $s=d+e-a-b-c$. We employ these relations to obtain Eq.\,(\ref{separateB5}) from Eq.\,(\ref{eq:B5}) by substituting Eq.\,(\ref{eq:HYPrelation1}) and then using Eq.\,(\ref{eq:HYPrelation2}).
Another relation that we use is the Stirling's formula for $\Gamma (z)$ function in the $|z|\rightarrow \infty$ and $|\arg\,z|<\pi$ limit:
\bea\label{Stirlingformula}
\Gamma(z)\rightarrow \sqrt{2\pi}e^{-z}z^{z-\frac{1}{2}},
\eea
for pole isolation in the $B_4$ and $B_5$ models.

\bibliography{DoubleReggeofB5model.bib} 

\end{document}